\documentclass[fleqn,usenatbib]{mnras}

\usepackage{lsstdesc_macros}
\usepackage{subfig}
\usepackage{graphicx}
\usepackage{threeparttable}
\usepackage{tcolorbox}
\usepackage[ruled,vlined]{algorithm2e}
\usepackage{amsmath,amssymb,bm}
\usepackage{tabularx}
\usepackage{orcidlink}
\usepackage{bm}
\usepackage[T1]{fontenc}
\usepackage{aecompl}

\usepackage{hyperref}
\graphicspath{{./}{./figures/}}
\newcommand{\refresponse}[1]{{\textcolor{black}{#1}}}

\newcommand{\ve}[1]{\vect{#1}}
\newcommand{\mt}[1]{\bm{\mathsf{#1}}}

\newcommand{\nz}{\vect{\phi}_{\rm nz}}
\newcommand{\nzs}[1]{\vect{\phi}_{\rm nz#1}}

\title[Photo-z error in shear analysis]{Photometric Redshift Uncertainties in Weak Gravitational Lensing Shear Analysis: Models and Marginalization}

\author[]{Tianqing Zhang$^1$\orcidlink{0000-0002-5596-198X}, Markus Michael Rau$^{1,2}$\orcidlink{https://orcid.org/0000-0003-3709-1324}, Rachel Mandelbaum$^1$\orcidlink{0000-0003-2271-1527}, Xiangchong Li$^1$, \newauthor  Ben Moews$^{1,3}$\orcidlink{0000-0003-0897-040X}
 \\
$^1$McWilliams Center for Cosmology, Department of Physics, Carnegie Mellon University, 5000 Forbes Ave, Pittsburgh, PA 15213, USA.
 \\
 $^{2}$High Energy Physics Division, Argonne National Laboratory, Lemont, IL 60439, USA\\
$^3$Pittsburgh Supercomputing Center, Carnegie Mellon University \& University of Pittsburgh, 300 S Craig St, Pittsburgh, PA 15213, USA.
}

\date{\today}

\begin{document}

\label{firstpage}
\pagerange{\pageref{firstpage}--\pageref{LastPage}}
\maketitle

\begin{abstract}
Recovering credible cosmological parameter constraints in a weak lensing shear analysis requires an accurate model that can be used to marginalize over nuisance parameters describing potential sources of systematic uncertainty, such as the uncertainties on the sample redshift distribution $n(z)$. Due to the challenge of running Markov Chain Monte-Carlo (MCMC) in the high dimensional parameter spaces in which the $n(z)$ uncertainties may be parameterized, it is common practice to simplify the $n(z)$ parameterization or combine MCMC chains that each have a fixed $n(z)$ resampled from the $n(z)$ uncertainties.  In this work, we propose a statistically-principled Bayesian resampling approach for marginalizing over the $n(z)$ uncertainty using multiple MCMC chains. We self-consistently compare the new method to existing ones from the literature in the context of a forecasted cosmic shear analysis for the HSC three-year shape catalog, and find that these methods recover \refresponse{statistically consistent errorbars for the} cosmological parameter constraints \refresponse{for predicted HSC three-year analysis}, implying that using the most computationally efficient of the approaches is appropriate. However, we find that for datasets with the constraining power of the full HSC survey dataset (and, by implication, those upcoming surveys with even tighter constraints), the choice of method for marginalizing over $n(z)$ uncertainty among the several methods from the literature may modify the \refresponse{$1\sigma$} uncertainties on \refresponse{$\Omega_m-S_8$ constraints by $\sim$4\%,} and a careful model selection is needed to ensure credible parameter intervals. 

\end{abstract}


\section{Introduction}
\label{sec:int:0}

Over the past decade, wide-field imaging surveys, e.g., the Dark Energy Survey \citep[DES;][]{des_review}, the Kilo-Degree Survey \citep[KiDS;][]{deJong:2017bkf}, and the Hyper Suprime-Cam Subaru Strategic Program  \citep[HSC SSP;][]{2018PASJ...70S...4A}, became increasingly powerful, reaching fainter magnitudes and larger areas, and employing improved methods for  controlling systematic biases and uncertainties \citep[for a review, see ][]{Mandelbaum:2017jpr}. Future surveys such as the Vera C.\ Rubin Observatory Legacy Survey of Space and Time \citep[LSST;][]{Ivezic:2008fe, 2009arXiv0912.0201L}, the \textit{Nancy Grace Roman} Space Telescope High Latitude Imaging Survey \citep{2015arXiv150303757S, 2019arXiv190205569A} and \textit{Euclid} \citep{Euclid_overview} will provide even larger data volumes and require more stringent control of systematic errors.  With these developments, cosmic shear, the coherent weak gravitational lensing effect on the light from the distant galaxies caused by the large scale structure, becomes one of the most powerful probes to test the standard model of cosmology \refresponse{\citep{Hu:2001fb, 2010GReGr..42.2177H, 2020PASJ...72...16H, 2021A&A...645A.104A, 2022PhRvD.105b3514A,2022PhRvD.105b3515S}}.

The prevalent method of cosmological parameter analysis based on cosmic shear currently relies on tomographic binning \citep{1999ApJ...522L..21H} and measuring the two-point correlation function (2PCF) of the source galaxy shapes \refresponse{\citep[e.g.,][]{Hildebrandt:2018yau, 2020PASJ...72...16H,2022PhRvD.105b3514A,2022PhRvD.105b3515S}}. For this approach to cosmological analysis, the distribution of the source galaxy distances along the line-of-sight, commonly known as the sample redshift distribution $n(z)$, is an important quantity for forward modeling the auto- or cross-2PCF of cosmic shear within or between tomographic bins, respectively \citep[e.g.,][]{2006MNRAS.366..101H}. 

Due to the expense of spectroscopic observations for galaxy samples at the depths of current imaging surveys, weak lensing measurements typically rely on multi-band photometric redshifts as their initial source of redshift information, having only limited and typically not representative training samples with spectroscopic redshifts. There two primary categories of  photometric redshift estimation methods \citep[for a review, see][]{2019NatAs...3..212S} are as follows: (a) template fitting, which is based on finding the best-fit spectral energy distributions (SED) template by fitting to the broad-band photometry; and (b) machine learning methods, which use the training sample to learn a relationship between redshift, photometry,  and potentially other information, e.g., morphological parameters. The outputs of these photo-z methods are normally probability density functions for individual galaxies, which we will call $p(z)$.

Deriving the aforementioned sample redshift distributions $n(z)$ based on uncertain and potentially biased individual galaxy $p(z)$ is highly non-trivial \citep[e.g.,][]{2021PhRvD.103h3502M}, as doing so properly requires deconvolution of the uncertainties and correction for any biases. 
Methods for reconstructing properly calibrated $n(z)$ include  direct calibration based on magnitude re-weighting to match a reference sample with known redshifts \citep[DIR;][]{2009MNRAS.396.2379C} and cross-correlating spectroscopic samples and photometric samples \citep[CC;][]{2008ApJ...684...88N,2019MNRAS.483.2801S,2020MNRAS.491.4768R}. Additionally, some methods aim to estimate $n(z)$ directly from photometric observables instead of using  photometric redshifts~\citep[see, e.g.,][]{Lima_2008}, with the latter branching into machine learning and related approaches in recent years~\citep{Malz_2018, Henghes_2021}. More recent work permits the combination of the $p(z)$ with a regularized deconvolution of their uncertainty, in combination with the CC method \citep{2022MNRAS.509.4886R} -- a method that is being applied in practice by Rau et al., {\em in prep.} to data from the HSC survey. 

Since the cosmic shear signal is sensitive to the sample redshift distribution, it is necessary to carefully model the uncertainties on $n(z)$ and marginalize over them for the current and upcoming surveys \citep{Malz_2022}. The marginalization is not a trivial task, since the uncertainties on $n(z)$ are often modeled in a high dimensional space, making attempts to run a full MCMC extremely computationally intensive. 
Therefore, several methods have been used to  approximately marginalize over the redshift distribution uncertainties.  This includes allowing just a shift in the mean redshift of the $n(z)$ for each tomographic bin, a method that has been adopted in many cosmology analyses \refresponse{\citep[e.g.,][]{2020PASJ...72...16H, Joudaki:2019pmv,2022PhRvD.105b3514A,2022PhRvD.105b3515S}}. 
In other cases, methods have been developed to marginalize over realistic uncertainties on $n(z)$, for example by (a) combining 750 MCMC chains each run with a different random realization sampled from the prior for $n(z)$  \citep{2017MNRAS.465.1454H}, (b) analytically approximating the likelihood function on the redshift nuisance parameters \citep{Hadzhiyska_2020, 2021A&A...650A.148S}, and (c) ranking $n(z)$ realizations in a lower dimensionality latent space to reduce the number of nuisance parameters \citep{2022MNRAS.511.2170C}. 

In this study, we develop and apply methodology to systematically compare the performance of methods of $n(z)$ uncertainty marginalization for cosmic shear.  Our goal is to quantify tradeoffs such as systematic bias, credible uncertainty estimation, and computational costs.  For this purpose, we start by presenting the new resampling approaches for marginalizing over uncertainties in the sample redshift distribution $n(z)$.  We apply the new method and compare it with several existing approaches in the literature, in the context of cosmic shear with the three-year HSC shear catalog \citep[HSC Y3;][]{2021arXiv210700136L}.  We consider the above-mentioned tradeoffs and make a recommendation for methodology that would be appropriate for cosmology analysis of the HSC Y3 shear catalog. 

The structure of this paper is as follows. In Section~\ref{sec:bgd:0}, we provide  brief background on $n(z)$ uncertainty modeling and the tomographic 2PCF cosmological analysis of cosmic shear. In Section~\ref{sec:mtd:0}, we outline the approaches we will explore for marginalization over ensemble redshift uncertainties, including the new method and several pre-existing methods in the literature. We also explain the specific setup for the cosmological analysis we use for comparing these methods. In Section~\ref{sec:res:0}, we show the results for the cosmological parameter inference using multiple approaches for redshift uncertainty marginalization. In Section~\ref{sec:conclu}, we summarize our findings in this paper and discuss their practical implications.

\section{Background}
\label{sec:bgd:0}

In this section, we provide the background that motivates this study. In Section~\ref{sec:bgd:wl_analysis}, we introduce the weak lensing shear analysis paradigm of this paper, and describe the modeling and marginalization of redshift distribution uncertainties in previous shear analyses. Section~\ref{sec:bgd:nz_estimation} describes our flexible parametrization for the sample redshift distribution and discusses our choice of prior on the associated sample redshift distribution model parameters. 

\subsection{Weak Lensing Shear Analysis}
\label{sec:bgd:wl_analysis}

In this work, we  discuss marginalization over the $n(z)$ uncertainties in a tomographic weak lensing shear analysis \citep{1999ApJ...522L..21H} based on the two-point correlation function \citep[2PCF;e.g.,][]{1980lssu.book.....P,2000astro.ph..3338K,2008A&A...479....9F,2014MNRAS.440.1322H}. In this section, we provide a brief background of this analysis paradigm. We define terms, e.g., the data vector (observable) and its covariance matrix, and the forward model that predicts the theoretical value of the observable given cosmological and nuisance parameters. Among nuisance parameters, we emphasize the parameterization of the redshift distribution uncertainties, which is the focus of this paper. 

The goal of the weak lensing shear analysis is to extract information about the cosmological model from the shear 2PCF. The cosmic shear observable that is commonly measured in real space analyses \refresponse{\citep[e.g.,][]{2020PASJ...72...16H, Joudaki:2019pmv, 2022PhRvD.105b3514A,2022PhRvD.105b3515S}} 
is the correlation functions of the observed galaxy shears $\xi^{ij}_\pm(\theta)$, where $\theta$ is the angular separation of the galaxies, and $i$ and $j$ are the indices of the tomographic bin pair. The data vector is obtained by concatenating $\xi^{ij}_\pm(\theta)$ from different tomographic bin pairs across all angular bins used for the measurement. 

The observed data vector $\vect{D}$ is compared to the theoretical data vector $\vect{T}$, which is predicted by a forward modeling pipeline that considers the cosmological parameters and systematic biases and uncertainties, e.g., the uncertainties on the redshift distribution, and the intrinsic alignment of galaxy shapes due to gravitational tidal effects \citep[IA;][]{2000ApJ...545..561C, 2000MNRAS.319..649H}. The log-likelihood of a model parameter vector $\ve{\Omega}$ is computed by 
\begin{equation}
    \label{eq:likelihood_define}
    \log(\mathcal{L}(\ve{\Omega} | \vect{D})) = (\vect{D}-\vect{T}(\ve{\Omega})) \mt{\Sigma}^{-1} (\vect{D}-\vect{T}(\ve{\Omega}))^T,
\end{equation}
where $\mt{\Sigma}$ is the covariance matrix of $\vect{D}$. 
MCMC samplers such as \textsc{MultiNest} \citep{Feroz_2008, 2009MNRAS.398.1601F, Feroz_2019} are used to efficiently sample over the parameter space and provide parameter inferences based on the likelihood in Eq.~\eqref{eq:likelihood_define} and the prior information on the parameters. 

An important step to forward model the shear-shear 2PCF in tomographic bin pairs is to project the 3-D matter power spectrum $P(k,z)$ to the 2-D angular shear power spectrum $C_\ell^{ij}$. Under the Limber approximation, the angular shear power spectrum \refresponse{\citep{1998ApJ...506...64S, 1999ApJ...522L..21H}} between bins $i$ and $j$ is
\begin{equation}
    \label{eq:angular_power}
    C_\ell^{ij} = \int \frac{\mathrm{d}\chi}{\chi^2} P(\ell/\chi;z(\chi)) q^i(\chi) q^j(\chi), 
\end{equation}
where $P(\ell/\chi;z(\chi))$ is the matter power spectrum at $z$.  
$q^i(\chi)$ and $q^j(\chi)$ are the corresponding lensing efficiency function for tomographic bins $i$ and $j$. $q^i(\chi)$ is directly determined by the underlying redshift distribution $n^i(z)$: 
\begin{equation}
    \label{eq:transfer_function}
    q^i(\chi) = \frac{3\Omega_m H_0^2}{2c^2}\frac{\chi}{a(\chi)} \int_\chi^{\chi_h} \mathrm{d}\chi' n^i(\chi'(z))\frac{\chi' - \chi}{\chi'},
\end{equation}
where  $\Omega_m$ is the matter density parameter, $H_0$ is the Hubble constant, $\chi$ is the comoving radial distance, $a$ is the scale factor, and $c$ is the speed of light \refresponse{\citep[e.g.,][]{2015RPPh...78h6901K,2017MNRAS.470.2100K}}. \refresponse{Here we have used the formalism for a flat geometry.} We can see that $n^i(z)$ is a key factor determining the angular shear power spectrum, which itself  directly determines the shear-shear 2PCF $\xi_\pm^{ij}$ \refresponse{\citep{2001PhR...340..291B,2010A&A...523A...1J}}. Under the flat-sky approximation, $\xi_\pm^{ij}$ is expressed as
\begin{equation}
    \label{eq:shear_shear_2pcf}
    \xi_\pm^{ij}(\theta) = \frac{1}{2\pi} \int \mathrm{d}\ell \ell C_\ell^{ij}  J_{2\mp2} (\ell \theta),
\end{equation}
where $J_n$ is the n-th order Bessel function of the first kind.  
This deep connection between the redshift distribution and the cosmic shear observables is the reason why it is important to marginalize over the uncertainties on $n(z)$ to recover credible cosmological parameter constraints.

The sample redshift distribution $n(z)$ is often modeled as arrays of histogram bin heights $\nz$, as is further described in Sec.~\ref{sec:bgd:nz_estimation}. Since sampling in high dimensional parameter spaces is very computationally expensive, it may not be possible to model the sample redshift distribution uncertainties in every redshift bin that $n(z)$ is estimated on. 
A majority of previous shear analysis \refresponse{\citep[e.g.,][]{2020PASJ...72...16H, Joudaki:2019pmv, 2022PhRvD.105b3514A,2022PhRvD.105b3515S}} parameterized the redshift distribution of bin $i$ by allowing its mean redshift to shift,
\begin{equation}
    \label{eq:shifting_model}
    n^i(z) = n^i(z - \Delta z^i).
\end{equation}
With the shift model, the number of free parameters is equal to the number of the tomographic bins. The priors on these parameters are determined by the prior distributions of the calibrated $n(z)$. The shift model tremendously reduces the number of parameters compared to use of all histogram bin heights $\nz$,  
though it suffers from a limited number of degrees of freedom compared to the realistic $n(z)$ uncertainties. With cosmic shear analysis becoming increasingly systematics-dominated as the statistical uncertainties become smaller, 
various methods have been introduced to marginalize over a more realistic estimate of the $n(z)$ prior. In \cite{Hildebrandt:2018yau}, 750 realizations were drawn from the $n(z)$ prior, after which cosmic shear analyses were run on each realization. The chains were then directly concatenated to derive constraints on the cosmological parameters, including their uncertainties. \cite{2021A&A...650A.148S} applied the Laplace approximation to the prior of the redshift parameters and assumed the likelihood function is a multivariate Gaussian, thereby analytically marginalizing over the redshift parameter using the self-calibration algorithm. In \cite{2022MNRAS.511.2170C}, realizations of $n(z)$ were drawn from the prior distribution, then mapped into a lower-dimensional latent space, within which the likelihood function is smooth. 

In this paper, we revisit some of the methods mentioned above to marginalize over the $n(z)$ uncertainties, carrying out tests on mock cosmic shear analyses. 
We propose a new method of marginalizing over the $n(z)$ uncertainties based on statistical principles. By comparing the new method to other options, we aim to provide the optimal approach for the HSC Y3 cosmic shear analysis.

\subsection{Prior Specification on the Sample Redshift Distribution}
\label{sec:bgd:nz_estimation}


In this section, we briefly summarize how a prior on the sample redshift distribution was specified. For a discussion on the $n(z)$ inference methodology we refer to Rau et al. (in prep.). 

As shown in Eq.~\eqref{eq:angular_power}, the sample redshift distribution enters the modelling of two point functions via the transfer function in Eq.~\eqref{eq:transfer_function}. The entire redshift range is subdivided into $N_{\rm bins}$ histogram bins, and  
the sample redshift distribution in the $i$-th tomographic bin is parametrized as
\begin{equation}
    n^i(z; \nz^i) = \sum_{k = 1}^{N_{\rm bins}} \nzs{,k}^i \vect{1}(z \in [z_L^{k}, z_R^{k}]) \, , 
    \label{eq:nz_definition}
\end{equation}
where $[z_L^{k}, z_R^{k}]$ denotes the left/right edges of histogram bin $k$. $\nzs{,k}^i$ is the $k$-th histogram bin height in the $i$-th tomographic bin. $\vect{1}$ is the indicator function. The distinction between `histogram bin' and tomographic bin is as follows: the former denotes the bins of the histogram parametrization, the latter denotes the selection bins of the tomography. Eq.~\ref{eq:nz_definition} defines the histogram heights vector $\nz^i$ as the parameters of a linear basis function model for the sample redshift distribution with tophat basis functions. 

The prior $p(\nz^i)$, i.e., uncertainties on the sample redshift distribution histogram bin heights in the $i$-th tomographic bin, is inferred using an extension of the methodology developed in \citet{2022MNRAS.509.4886R}. It combines information from both spatial cross-correlations of a reference sample with spectroscopic redshifts and a sample with photometric redshift information. We reiterate that a future publication will provide more details of the inference methodology (Rau, et al, {\em in prep.}). The method utilizes the `S16A CAMIRA-LRG sample' \citep{2021ApJ...922...23I}, a sample of Luminous Red Galaxies selected using the CAMIRA algorithm \citep{2014MNRAS.444..147O} from the HSC data observed in the first observing season of 2016, as a reference sample.  This choice can be motivated by the accurate photometric redshift estimates that are available for the LRGs (relative to the photometric redshift errors in the full HSC S16A sample), and a sufficiently high number density. 

The spatial cross-correlation between the CAMIRA-LRG 
sample (c) and a photometric sample (p) can be predicted as 
\begin{equation}
    \vect{w_{\rm sc}} \propto \nzs{,p} \, \vect{b_p} \,  \vect{b_c}  \, \vect{w_{\rm DM}} \, , 
    \label{eq:wx_definition}
\end{equation}
where $\nzs{,p}$ denotes the parameters of the sample redshift distribution,  
($\vect{b_p}$/$\vect{b_c}$) denote the galaxy-dark matter bias parameters of the (photometric/CAMIRA-LRG) samples in each redshift bin and $\vect{w_{\rm DM}}$ denotes the dark-matter contribution to the cross-correlation signal. We present a simplified vector notation, where the elements in Eq.~\ref{eq:wx_definition} correspond to the cross-correlation measurements in each redshift bin, obtained by measuring the correlation amplitude within a spatial annulus of physical distance as described in \citet{2017MNRAS.467.3576M}. Using the auto-correlation of the CAMIRA-LRG galaxies the method fits the linear bias model $b_c(z) = b_0(1+z)$, where $b_0 = 1.06 \pm 0.03$, consistent with previous measurements from \citet{2021arXiv210308628I}. The covariance of the cross-correlation likelihoods is estimated using bootstrap resampling and approximated to be diagonal. 
This is done for simplicity and can be an inaccurate approximation due to the high correlation of neighboring bins. The method uses \textsc{the-wizz}\footnote{\url{https://github.com/morriscb/The-wiZZ/}} \citep{2017MNRAS.467.3576M} for the cross-correlation measurements, and selects a scale annulus of $1.5-5.0 \, {\rm Mpc}$ in analogy to \citet{2020arXiv201208569G}.

We include information from the photometry into the inference by combining the individual galaxy redshift uncertainties of a set of models. Our model set consists of a template fitting code \textsc{Mizuki} \citep{2015ApJ...801...20T} that defines a likelihood, empirical codes \textsc{MLZ}\footnote{\url{https://github.com/mgckind/MLZ}} \citep{2013MNRAS.432.1483C} and \textsc{EPHOR} \citep{2018PASJ...70S...9T} that define a conditional probability density function obtained on a training set and  \textsc{Franken-Z}\footnote{\url{https://github.com/joshspeagle/frankenz}} \citep{2019MNRAS.490.5658S} that uses a flux-error weighted score function to map training set objects to galaxies in the photometric dataset. We refer to \citet{2018PASJ...70S...9T} for a summary of the different methodologies that are available to us. We note that the machine learning-based algorithms do not produce likelihoods (unlike SED fitting techniques). However we will treat their estimates as likelihoods within this framework and refer to a future publication for a description of the technical details. 

\begin{figure}
    \centering
    \includegraphics[width=1\columnwidth]{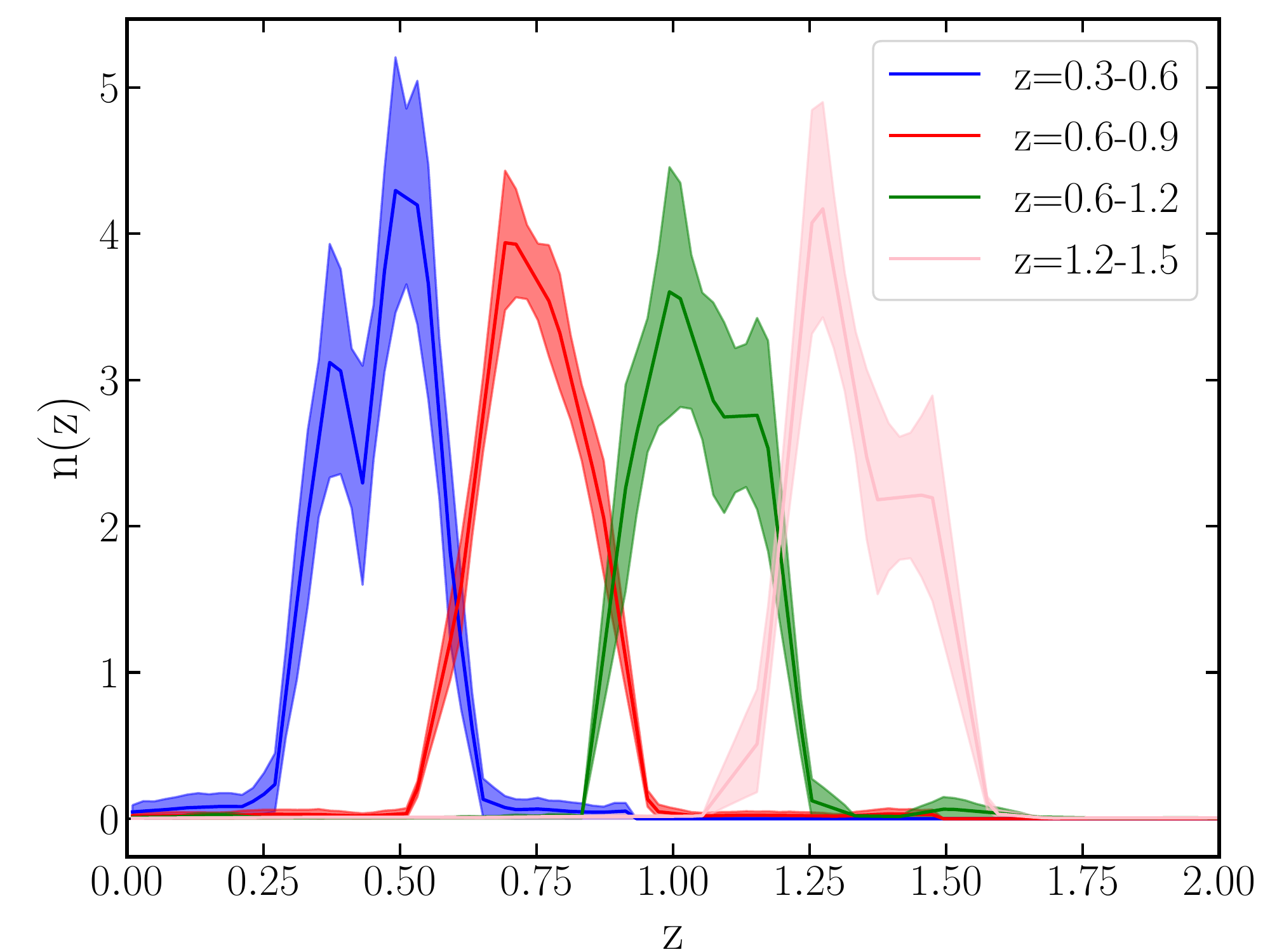}
    \caption{The sample redshift distribution estimated by cross correlation with 4 tomographic bins, for HSC S16A (Rau, et al., {\em in prep.}) The shaded regions represent the $68\%$ confidence intervals of the distributions. 
    }
    \label{fig:nz_cc}
\end{figure}

Following the methodology developed in \citet{2022MNRAS.509.4886R}, we infer posteriors of sample redshift distributions as shown in Fig.~\ref{fig:nz_cc} using information from both the cross-correlation data vector and the photometry of galaxies. The horizontal axis shows the redshift, the vertical the normalized sample redshift distribution. The legend lists the redshift ranges selected on the best fitting redshift derived using the \textsc{Mizuki} template fitting code that we use to define the tomographic bins. The error contours correspond to the $68\%$ confidence intervals. The aforementioned posteriors of sample redshift distributions constructed using the joint likelihood of spatial cross-correlations and photometry is then used as the prior distribution on the sample redshift distribution in the following analysis. We neglect here the covariance between the spatial cross-correlations and the lensing observables.

\refresponse{In this work, we assume that the uncertainties in the ensemble redshift distribution for the HSC three-year and full analysis do not significantly decrease compared with those for the first-year HSC analysis. Constraints on the sample redshift distribution are limited by (a) practical issues such as the redshift range of the LRG sample and our knowledge of the galaxy-dark matter bias; (b) the model uncertainty between photometric redshift codes, estimated using the COSMOS2015 field \citep{2016ApJS..224...24L}. The modeling uncertainty is limited by the cosmic variance, and is independent of the survey area, therefore will not decrease for the HSC three-year analysis compared with the first-year analysis. As a result, the redshift uncertainties are expected to decrease much more slowly than the cosmic shear covariance matrix as the survey area grows.}

\begin{table*}
\begin{tabularx}{\textwidth}{llX}
\hline
Terminology    &  Symbol       & Description  \\\hline
$n^i(z)$ prior   & $P(\nz^i|\vect{\alpha}^i)$           & The prior on the $n^i(z)$ histogram bin heights in the $i$-th tomographic bin. Specifically, we adapt the posterior in Rau et al.\ (in prep.) $P(\nz^i|\vect{\alpha}^i)$ parameterized on the Dirichlet parameter $\vect{\alpha}^i$ for the $i$-th tomographic bin, as the prior, which is described in Section~\ref{sec:bgd:nz_estimation}. We sometimes refer to this as $n(z)$ prior, when the tomographic bin is not specified. \\\\

Average $n^i(z)$ & $\langle \nz^i \rangle$ & The average histogram bin heights for $n^i(z)$ in the $i$-th tomographic bin, averaged over 10000 realizations of $\nz^i$ sampled from the $n(z)$ prior. \\\\

Mean redshift  & $\langle z^i \rangle$      & The mean redshift of the $i$-th tomographic bin, calculated by $\langle z^i \rangle = \int z P(\nz^i|\vect{\alpha^i}) \mathrm{d}z$, 
where $P(\nz^i|\vect{\alpha}^i)$ 
is the $n^i(z)$ prior of the samples in the $i$-th tomographic bin. \\\\

Data vector  & $\vect{D}$      & Shear data vector $\vect{D} = [\vect{\xi^{ij}_+}, \vect{\xi^{ij}_-}]$, where $ij$ is ordered in $[11,12,13,14,22,23,24,33,34,44]$. The generation of data vector is described in Section~\ref{sec:mtd:mock}. \\\\

Covariance (matrix)  & $\vect{\Sigma}$      & Covariance matrix of the data vector $\vect{D}$, $\Sigma_{ij} = \langle D_i D_j \rangle$. The covariance matrix used in this work is described in Section~\ref{sec:mtd:2pt_like}. \\\\

Inference posterior & $P(\vect{\Omega} | \vect{D})$  & The posterior distribution on the cosmological and astrophysical parameters $\vect{\Omega}$ after marginalizing over the nuisance parameters. In this paper, we specifically consider the $n(z)$ parameters as the nuisance parameters. \\\\

Log evidence & $\log(P(\vect{D}|\nzs{,s}))$ & The log-evidence of a particular realization of the $\nzs{,s}$, expressed in Eq.~\eqref{eq:bayesian_evidence}.  \\\\

Number of tomographic bins & $N_{\rm tomo}$ & The number of tomographic bins, which results in the number of nuisance parameters for the multiplicative bias and shift model. In this work, $N_{\rm tomo} = 4$\\\\

Number of resampling $\nz$ & $N_{\rm sample}$ & The number of realizations sampled from the $n(z)$ prior for the direct and Bayesian resampling methods, described in Sec.~\ref{sec:mtd:resample}.  For the full analyses in this work, $N_{\rm sample}=250$.\\\\

Number of histogram bins & $N^i_{\rm bins}$ & Number of histogram bin heights in the $i$-th tomographic bins. This is the same as the length of $\nz^i$. In this work, $N^i_{\rm bins} = 18(18,25,20)$ for tomographic bins 1(2,3,4), respectively.\\

\hline
\end{tabularx}
\caption{Table of the redshift distribution and statistics terminologies used for quantities throughout Section~\ref{sec:mtd:0}. We also provide a short description of each quantity and the specific values used in this work or a reference to the section where they are described. 
}
\label{tab:terminology}
\end{table*}

\section{Methods}
\label{sec:mtd:0}

In this section, we describe the methods used to carry out this work.  In Section~\ref{sec:mtd:pipeline}, we describe our parameter inference pipeline, implemented using CosmoSIS \citep{2015A&C....12...45Z}. 
In Section~\ref{sec:mtd:marginal}, we describe the methods to marginalize over the $n(z)$ uncertainties during the cosmological parameter inference. In addition to employing existing approaches from the literature, we also propose a new method for marginalizing $n(z)$ uncertainties for cosmic shear analysis: a statistically accurate formulation for sampling from the $n(z)$ covariance. 

The key terminology used for redshift distribution and statistical inference throughout this section, their mathematical symbols, and description are listed in Table~\ref{tab:terminology}. 

\subsection{Cosmological forward modeling}
\label{sec:mtd:pipeline}

In this section, we describe the cosmic shear forward modeling process, including the cosmological model, the astrophysical model, and other nuisance parameters, for computing the mock data vector and parameter inference. For an initial exploration, we considered a 2-parameter $\Lambda$CDM model that only varies $\Omega_m$ and $\sigma_8$. We then considered a full analysis with 5 $\Lambda$CDM parameters, 2 astrophysical nuisance parameters, $N_{\rm tomo} = 4$ multiplicative bias parameters and 2 PSF systematics parameters, for a total of 13 parameters. $N_{\rm tomo}=4$ additional parameters were added for marginalizing over $n(z)$ uncertainties for the shift model. 
The modeling pipeline used CosmoSIS \citep{2015A&C....12...45Z}, which is a well-tested and validated platform for cosmological inference \citep[e.g., in][]{2022PhRvD.105b3520A}.

The cosmological model is described in Section~\ref{sec:mtd:cosmology}, while the astrophysical and other nuisance parameters are described in Section~\ref{sec:mtd:nuisance}. The analysis setup (tomographic bins, angular scales, etc.) and mock data vector are described in Section~\ref{sec:mtd:mock}.  The sampler and covariance matrices are described in Section~\ref{sec:mtd:2pt_like}. 

\subsubsection{Cosmological Model}
\label{sec:mtd:cosmology}

We adopted a $\Lambda$CDM cosmological model throughout this work.  We computed the linear matter power spectrum using \textsc{CAMB} \citep{2000ApJ...538..473L, Lewis_2002, Howlett_2012}, and the nonlinear matter power spectrum using the updated \textsc{halofit} \citep{2012ApJ...761..152T} from the original version \citep{2003MNRAS.341.1311S}. The neutrino mass $\Omega_\nu$ was fixed to zero, since the weak lensing shear is relatively insensitive to it. 
The cosmological parameters in our model are provided in Table~\ref{tab:cosmology_parameters}, including their fiducial values, priors, and whether they are varied or fixed in our analysis.

\subsubsection{Astrophysical and Nuisance Parameters}
\label{sec:mtd:nuisance}

Throughout the analysis, we used the nonlinear alignment model \citep[NLA][]{2017MNRAS.470.2100K}  to model the intrinsic alignment (IA) signal~\citep[see also][for the development and further extension of the NLA model]{Hirata_2004, Bridle_2007}. 
In this paper, we adopted the NLA model with an additional term that includes redshift evolution of the alignment amplitude, namely,
\begin{equation}
    \label{eq:ia_amplitude}
    A(z) = A_{\text{IA}} \left[ \frac{1+z}{1+z_0} \right]^\eta,
\end{equation}
where the fiducial values and priors of the parameters $A_{\text{IA}}$, $\eta$, and $z_0$ are shown in Table~\ref{tab:nuisance_parameters}. In practice, the redshift evolution parameter may absorb some evolution of the source sample properties with redshift, since intrinsic alignments depend on galaxy properties.

Since the IA model in this work has redshift evolution, the intrinsic alignment model parameters may have some degeneracy with the redshift distribution $n(z)$, which motivates marginalizing over the $n(z)$ uncertainty in the analysis.  

We computed the shear-shear angular power spectrum from the matter power spectrum and the input $n(z)$, using the formalism in Section~\ref{sec:bgd:wl_analysis}. We then added the NLA model shear-IA and IA-IA angular power spectrum to the shear-shear angular power spectrum. Next, we included a per-bin multiplicative shear bias into the observed shear power spectrum using
\begin{equation}
\label{eq:multiplicative_bias}
\hat{C}_\ell^{ij} = (1 + m^{i} + m^j )C_\ell^{ij},
\end{equation}
where $m^i$ and $m^j$ are the multiplicative biases of bins $i$ and $j$, respectively. We used Eq.~\eqref{eq:shear_shear_2pcf} to compute the shear-shear correlation function $\xi_{\pm}^{ij}$. 

Finally, we employed a simple model for the additive shear biases at the correlation function level. We included the PSF leakage term $\alpha$ and PSF shape error term $\beta$, using the same model as in \cite{2020PASJ...72...16H}.  
Our model of the shear-shear correlation function with PSF systematics is
\begin{align}
    \label{eq:psf_systematics}
    \nonumber \xi_{\pm}^{ij,\text{model}} = \xi_{\pm}^{ij} + & \alpha^2 \langle e_{\text{PSF}} e_{\text{PSF}} \rangle \\ &+ \alpha \beta \langle e_{\text{PSF}} \delta e_{\text{PSF}} \rangle + \beta^2 \langle \delta e_{\text{PSF}} \delta e_{\text{PSF}} \rangle,
\end{align}
where $e_{\text{PSF}}$ and $\delta e_{\text{PSF}}$ are the PSF shape and the modeling error of the PSF shape, respectively. 

Table~\ref{tab:nuisance_parameters} lists the astrophysical and other nuisance parameters, with their fiducial values, priors, and whether they are varied or fixed in our analysis.

\subsubsection{Analysis settings and mock data vector}
\label{sec:mtd:mock}

In this work, we used 4 tomographic bins, resulting in 10 tomographic bin pairs. We adopted the angular binning used in the real-space cosmic shear analysis of the first-year HSC catalog \citep{2020PASJ...72...16H}, i.e., 9 angular bins between $8.06$ arcmin and $50.89$ arcmin for $\xi_+$, and 8 angular bins between $32.11$ arcmin and $160.93$ arcmin for $\xi_-$.  Our data vector $\vect{D}$, which includes $\xi_+$ and $\xi_-$, has a length of 170.  

We generated mock data vectors using the forward modeling pipeline described above. To be able to compare the recovered parameter values with their true values, we did not add noise to the data vectors. 

We used the Planck results in \cite{2020A&A...641A...6P} for the fiducial cosmological parameters in Table~\ref{tab:cosmology_parameters}.  For the IA parameters in Table~\ref{tab:nuisance_parameters}, we adopted typical integer values for the amplitude \refresponse{$A_\text{IA}$} and redshift power $\eta$, and $z_0 = 0.62$ for the pivot redshift\footnote{\refresponse{We have used $z_0=0.62$ for consistency with previous analysis.  However, as described in \citet{2022arXiv220807179L}, this choice does not affect the results much; choosing the mean redshift for the HSC survey gives consistent results.}} \citep{2018PhRvD..98d3528T, 2020PASJ...72...16H}. 
We adopted the prior on $\alpha$ and $\beta$ from \cite{2020PASJ...72...16H}, and set the fiducial values to zero. 

Our mock shear data vector was generated by averaging the $\xi_\pm$ over 1000 realizations of $n(z)$ sampled from its prior. Note that \refresponse{the auto-correlation} $\xi^{ii}_\pm(\langle n(z) \rangle) \ne \langle \xi^{ii}_\pm(n(z) ) \rangle$\refresponse{, with up to $0.75\%$ difference, as is demonstrated in Appendix~\ref{sec:ap:nzave}. } 
Therefore, we cannot simply use the mean value of the $n(z)$ prior to generate the mock data vector.

\begin{table}\centering
\begin{tabular}{lllll}
\hline
Parameter            & Fiducial & Prior          & 2-p & full analysis \\ \hline
$\sigma_8$           & $0.824$    & $U[0.4,1.2]$     & \checkmark       & \checkmark       \\
$\Omega_b$           & $0.0489$ & $U[0.03, 0.07]$ &                  & \checkmark       \\
$n_s$                & $0.967$  & $U[0.87,1.07]$   &                  & \checkmark       \\
$h_0$                & $0.677$  & $U[0.55,0.9]$   &                  & \checkmark       \\
$\Omega_m$           & $0.311$  & $U[0.1, 0.8]$ & \checkmark       & \checkmark       \\
$\tau$               & $0.0561$   & const.         &                  &                  \\
$\Omega_\nu$         & $0.0$    & const.         &                  &                   \\
$w$                  & $-1.0$   & const.         &                  &                  \\
$w_a$                & $0.0$    & const.         &                  &                  \\\hline
\end{tabular}
\caption{Fiducial values and priors of the cosmological parameters used in this paper, along with whether or not they are varied (\checkmark) or not (blank) in the two-parameter (2-p) and full analysis. $U[a,b]$ represents a uniform distribution from $a$ to $b$.}
\label{tab:cosmology_parameters}
\end{table}

\begin{table}\centering
\begin{tabular}{lllll}
\hline
Parameter            & Fiducial & Prior          & 2-p & full analysis \\ \hline
$A_{\text{IA}}$                  & $1.0$  & $U[-5,5]$      &                  & \checkmark       \\
$\eta$             & $0.0$  & $U[-5,5]$      &                  & \checkmark       \\
$z_0$                & $0.62$   & const.         &                  &                  \\
\\
$m_1$                & $0.0$   & $\mathcal{N}(0,0.01)$&                  &      \checkmark            \\ 
$m_2$                & $0.0$   & $\mathcal{N}(0,0.01)$&                  &      \checkmark            \\ 
$m_3$                & $0.0$   & $\mathcal{N}(0,0.01)$&                  &      \checkmark            \\ 
$m_4$                & $0.0$   & $\mathcal{N}(0,0.01)$&                  &      \checkmark            \\ 
\\
$\alpha$                & $0.0$   & $\mathcal{N}(0,0.01)$&                  &     \checkmark             \\ 
$\beta$               & $0.0$   & $\mathcal{N}(0,1.11)$&                  &     \checkmark             \\ \hline
\end{tabular}
\caption{Fiducial values and priors of the astrophysical and nuisance parameters used in this paper, along with whether or not they are varied (\checkmark) or not (blank) in the two-parameter (2-p) and full analysis. $U[a,b]$ represents a uniform distribution from $a$ to $b$, while $\mathcal{N}(\mu,\sigma)$ represents a Gaussian distribution with mean value $\mu$ and standard deviation $\sigma$.}
\label{tab:nuisance_parameters}
\end{table}

\subsubsection{Sampler and Covariance Matrices }
\label{sec:mtd:2pt_like}

We sampled the parameter space and estimate the Bayesian evidence using \textsc{MultiNest} \citep{Feroz_2008, 2009MNRAS.398.1601F, Feroz_2019}, due to its rapid speed for relatively accurate evidence evaluation in constant efficiency mode\footnote{In \cite{2022arXiv220208233L}, it is shown that varying the efficiency can bias the model evidence for \textsc{MultiNest}, therefore, we fixed the efficiency of \textsc{MultiNest} to eliminate this bias and for its speed over \textsc{PolyChord} \citep{ Handley_2015, 2015MNRAS.453.4384H}}. 
We fixed the efficiency to 0.1, which is the default value for \textsc{MultiNest}, throughout this work. 
The log-likelihood of the model is computed by Eq~\eqref{eq:likelihood_define}, with the corresponding covariance matrices. 

In this work, we carried out our analyses with two covariance matrices: (a) We estimated the  covariance matrix for cosmic shear using the HSC three-year shear catalog.  For this purpose, we divided every element in the HSC first-year covariance $\mt{\Sigma}_{y1}$ \citep{2020PASJ...72...16H} by 3, since the survey area is roughly 3 times larger. We denote this covariance matrix as $\mt{\Sigma}_{y3} = \mt{\Sigma}_{y1}/3$. (b) We estimated the covariance matrix for cosmic shear with the full HSC survey, which is roughly 10 times the area of the first-year catalog. We denote this covariance as $\mt{\Sigma}_{\text{full}} =  \mt{\Sigma}_{y1}/10$. 
There are several significant limitations of this approximation to the future HSC analyses: (a) We decreased the covariance by a factor of the increase in survey area, without considering that the survey footprint has become considerably more contiguous, so the survey edge effects become less important. (b) We adopted the same angular binning and scale cuts 
as for the HSC first-year analysis, while those cuts are likely to be different for the upcoming three-year analysis and future analyses. 

However, we used the covariance matrix of the $n(z)$ prior from the first-year HSC shape catalog when analyzing the three-year and full data vector. In the real analyses, the covariance of the $n(z)$ for the three-year and full catalogs is likely to decrease.  However, it is a systematics-dominated quantity, so its uncertainty will not decrease with area as rapidly as does the cosmic shear data vector. Our choice to keep it fixed represents a conservative assumption regarding our ability to understand and control systematic biases and uncertainties in the photometric redshift estimation and the cross-correlation calibration of $n(z)$.  As a result of this choice, the impact of $n(z)$ uncertainty on the cosmological parameter constraints gets worse as the dataset grows.

\subsection{Marginalizing over $n(z)$ uncertainty}
\label{sec:mtd:marginal}

In this section, we introduce the different approaches for marginalizing over uncertainty in the ensemble $n(z)$ that are implemented on the mock cosmic shear analysis described in Section~\ref{sec:mtd:pipeline}.
In Section~\ref{sec:mtd:shift}, we introduce the shift model's parameterization. 
In Section~\ref{sec:mtd:resample}, we introduce the resampling approach, i.e., marginalizing over the sample redshift distribution uncertainties by running many chains with different realizations drawn from the $n(z)$ prior. We propose a new technique for weighting the chains when combining them, based on model evidence, motivated by Bayes theorem.

The $n(z)$ prior that is marginalized over in this work is specified by the histogram bin heights \refresponse{$\nzs{,k}^i$} at \refresponse{the} center redshift of the histogram $z_k$ for tomographic bin $i$.  respectively, modeled by 4 independent Dirichlet distributions. The Dirichlet distributions are parameterized by arrays $\vect{\alpha}^i$, with length  equal to the number of histogram bins in the corresponding tomographic bin, specified in Section~\ref{sec:bgd:nz_estimation}.

\subsubsection{Shift Model}
\label{sec:mtd:shift}

\begin{figure}
    \centering
    \includegraphics[width=1\columnwidth]{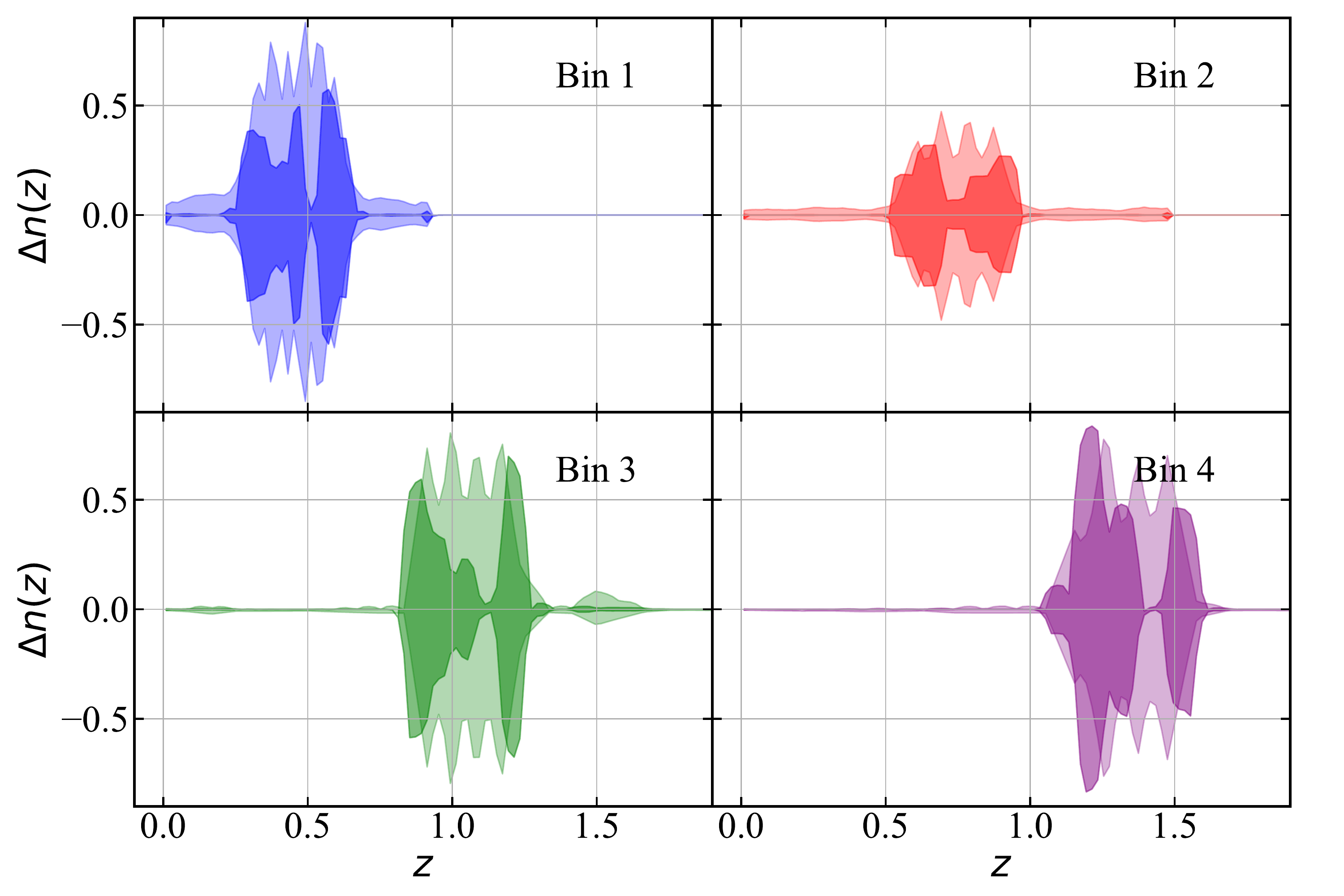}
    \caption{ The $68\%$ confidence intervals of the $n^i(z)$ uncertainties of the fiducial $n(z)$ prior (light shaded regions) and the shift model (dark shaded regions). The shift model generates an unrealistic distribution of $n(z)$ uncertainties, underestimating the uncertainty at most redshifts but overestimating it around the edges of bins 2, 3 and 4. }
    \label{fig:shift_vs_total}
\end{figure}

The shift model is a simple and approximate model for representing uncertainties in $n(z)$.  It allows the sample redshift distribution to shift coherently in redshift space following Eq.~\eqref{eq:shifting_model}. It is used to marginalize over $n(z)$ uncertainties in many cosmic shear analysis \refresponse{\citep[e.g.,][]{Hildebrandt:2018yau, 2020PASJ...72...16H,2022PhRvD.105b3514A,2022PhRvD.105b3515S}}. For this model, we use the average histogram bin heights $\langle \nz^i \rangle$ as the fiducial redshift distribution, specified in row~2 of Table~\ref{tab:terminology}. We let the $\langle \nz^i \rangle$ of each tomographic bin shift individually. Therefore, using this model involves introducing $N_{\rm tomo} = 4$ nuisance parameters. 
We determined the prior on the $\Delta z^i$ by computing the distribution of the mean redshift $\langle z^i \rangle$ of the tomographic bin $i$  
over 10000 realizations of histogram bin heights drawn from the $n^i(z)$ prior. We used a Gaussian distribution for the prior, with zero means and standard deviations determined by the distributions of $\langle z^i \rangle$. 
The priors on the shift parameters for the four tomographic bins are listed in Table~\ref{tab:shift_model_table}. 

In Fig.~\ref{fig:shift_vs_total}, we show a comparison of the $n(z)$ uncertainty included by the shift model (in dark shaded regions), versus the total uncertainty of the $n(z)$ prior (in light shaded regions). The uncertainty of the shift model is generated by shifting $\langle n^i(z) \rangle$ with $\Delta z^i$ sampled from the prior listed in Table~\ref{tab:shift_model_table}. Compared to the full $n(z)$ prior, the shift model underestimates the uncertainties at most redshifts, especially for redshifts where $\langle n^i(z) \rangle$ is relatively flat. The shift model also overestimates the uncertainties in the wings of the redshift distribution for some  the tomographic bins. At the $n(z)$ level, the shift model is an inaccurate representation of the real uncertainty. 

\begin{table}\centering
\begin{tabular}{lll}
\hline
Parameter    & Fiducial & Prior          \\\hline
$\Delta z^1$ & 0.0      & $\mathcal{N}(0,0.012)$ \\
$\Delta z^2$ & 0.0      & $\mathcal{N}(0,0.01)$ \\
$\Delta z^3$ & 0.0      & $\mathcal{N}(0,0.018)$ \\
$\Delta z^4$ & 0.0      & $\mathcal{N}(0,0.021)$ \\\hline
\end{tabular}
\caption{Fiducial values and priors used for the shift model parameterization of $n(z)$ uncertainties. The standard deviation of the Gaussian prior is calculated from the $\sigma$ of $\langle z^i \rangle$ from 1000 draws from the $n(z)$ prior.}
\label{tab:shift_model_table}
\end{table}

\subsubsection{Resampling Approaches}
\label{sec:mtd:resample}



A different approach for marginalizing over the $n(z)$ uncertainty, with fewer approximations, is to sample many realizations of histogram bin heights from the $n(z)$ prior, and run the cosmological parameter estimation process on each realization as if there are no $n(z)$ uncertainties. To incorporate the $n(z)$ uncertainties in the cosmological parameter estimates, the final step is to combine the results from the different MCMC chains. We refer to this approach as the ``resampling approach''.


In this work, we propose a resampling method that is based on Bayes' theorem to marginalize over the $n(z)$ uncertainty. We start by deriving the posterior on the cosmological and astrophysical parameters, $P(\ve{\Omega}|\vect{D})$, after marginalizing over the uncertainty in the $n(z)$ histogram bin heights, which we denoted $\nz$. Here $\vect{D}$ is the observed cosmic shear data vector $\vect{D} = [\vect{\xi}_+, \vect{\xi}_-]$, see row~4 of Table.~\ref{tab:terminology}.  This posterior is as follows:  
\begin{align}\nonumber
    P(\ve{\Omega}|\vect{D}) &= \int \mathrm{d} \nz P(\ve{\Omega}, \nz | \vect{D}) = \int \mathrm{d}\nz P(\ve{\Omega}|\vect{D}, \nz) P(\nz|\vect{D})\\
                &= \frac{1}{P(\vect{D})} \int \mathrm{d}\nz P(\ve{\Omega}|\vect{D},\nz) P(\vect{D}|\nz) P(\nz).
    \label{eq:resampling_integral_form}
\end{align}
The first line of the equation is based on conditional probability, and the second line is based on Bayes' theorem. 
Here $P(\ve{\Omega}|\vect{D},\nz)$ is the posterior on $\ve{\Omega}$ with a specific realisation of the redshift distribution $\nz$. $P(\nz)$ is the $n^i(z)$ prior, for which we chose to use $P(\nz|\vect{\alpha})$, the posterior probability distribution for the redshift distribution derived using an extension of the methodology from \citet{2022MNRAS.509.4886R}.  $P(\vect{D}|\nz)$ is the Bayesian evidence of the data given $\nz$, evaluated by integrating the joint conditional probability over $\ve{\Omega}$,
\begin{equation}
\label{eq:bayesian_evidence}
    P(\vect{D}|\nz) = \int \mathrm{d} \ve{\Omega} P(\vect{D}|\ve{\Omega},\nz) P(\refresponse{\ve{\Omega}}).
\end{equation}
We rely on the \textsc{MultiNest} estimation to the log-evidence, which is shown to have a constant bias from the truth in \cite{2022arXiv220208233L}, if the efficiency is kept fixed. This is fine for our purpose: the constant bias on the log-evidence results in a constant factor in the evidence, which is normalized out for the Bayesian weight $\omega_s$.

We now describe how we utilize the resampling approach to estimate $P(\ve{\Omega}|\vect{D})$ in Eq.~\eqref{eq:resampling_integral_form}.
We sampled $N_{\rm sample}$ realizations of the redshift distribution $\nzs{,s}$, where $s = 1 \dots N_{\rm sample}$, is the index of a particular realization  
from the $n(z)$ prior, i.e., $P(\nz|\vect{\alpha})$. We combined the inferred posterior distributions for each one (as represented by the MCMC chains), $P(\ve{\Omega}|\vect{D}, \nzs{,s})$. By doing so, we  effectively evaluated the integral of Eq.~\eqref{eq:resampling_integral_form}, which can be written the form of a summation, 
\begin{align}
\label{eq:resampling_summation}
P(\ve{\Omega}|\vect{D}) = \frac{1}{N_{\rm sample} P(D)} \sum_{s=1}^N P(\ve{\Omega}|\vect{D}, \nzs{,s} ) P(\vect{D}|\nzs{,s}),
\end{align}
where $\nzs{,s}$ is the $s$th sample of the redshift distribution.
Based on Eq.~\eqref{eq:resampling_summation}, we designed a Bayesian weight $\omega_s$ for combining the posteriors $P(\ve{\Omega}|\vect{D}, \nzs{,s} )$ that satisfies the following two conditions:
\begin{align}
    \omega_s & \propto P(\vect{D}|\nzs{,s})\\
    \sum_{s=1}^{N_{\rm sample}} \omega_s &= 1.
    \label{eq:weight_requirement}
\end{align}
Finally, the marginalized posterior of $\ve{\Omega}$ from the Bayesian resampling can be expressed as
\begin{equation}
P(\ve{\Omega}|\vect{D}) =  \sum_{s=1}^N P(\ve{\Omega}|\vect{D}, \nzs{,s} ) \omega_s.
    \label{eq:resampling_weights}
\end{equation}
Note that the constant $1/(N_{\rm sample} P(\vect{D}))$ in Eq.~\eqref{eq:resampling_integral_form} is absorbed in $\omega_s$ since summation of $\omega_s$ is normalized to 1. 
This weight $\omega_s$, which is proportional to the Bayesian evidence shown in Eq~\eqref{eq:bayesian_evidence}, preserves Bayes' theorem, effectively downweighting the $n(z)$ realizations that are not likely to generate the cosmic shear data vector $\vect{D}$. A similar resampling approach was used in \cite{Hildebrandt:2018yau}; however, the MCMC chains were concatenated with equal weights, which does not preserve  Bayes' theorem. We  therefore call our approach ``Bayesian resampling'', and call the method from \cite{Hildebrandt:2018yau} ``direct resampling'',  throughout the paper.

In principle, with enough samples of the redshift distribution, the Bayesian resampling approach should  accurately marginalize over the full prior on $n(z)$ in the cosmic shear analysis, giving more credible parameter constraints than simplified parameterizations, e.g., the shift model. However, it does have its drawbacks: (a) it is computationally intensive to run the full analysis for $N_{\rm sample}$ times, where $N_{\rm sample}$ is the number of redshift distribution samples, (b) it requires the sample redshift distribution $n(z)$ to have a well-defined probability distribution from which samples can be drawn, which might not be the case for some surveys depending on how they infer the ensemble $n(z)$.

\begin{figure*}
    \centering
    \includegraphics[width=1.0\columnwidth]{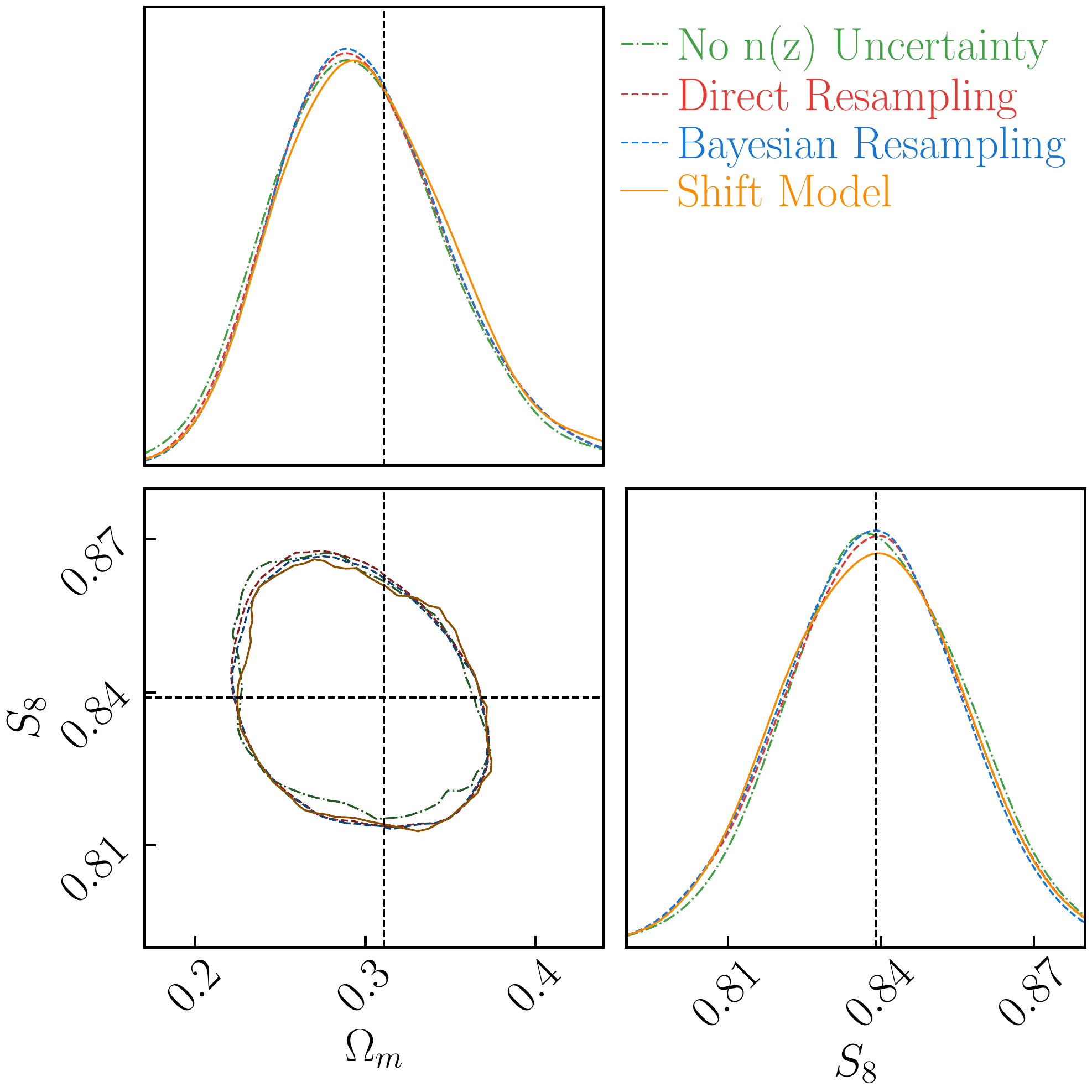}
    \includegraphics[width=1.0\columnwidth]{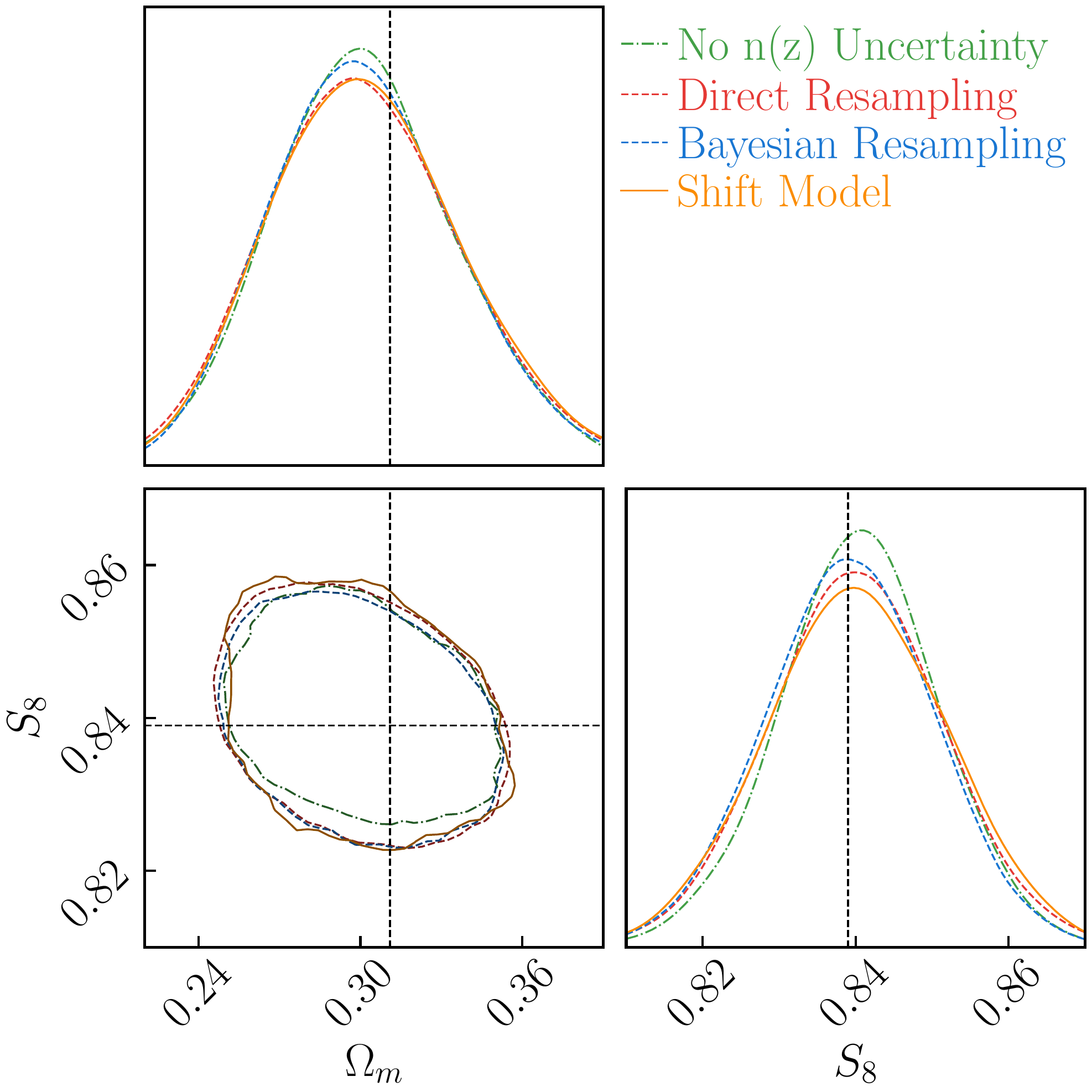}
    \caption{ Parameter constraints for the full analyses, with the three-year covariance matrix $\Sigma_{y3}$ (left), and the full-data covariance matrix $\Sigma_{\text{full}}$ (right), with parameters listed in Table~\ref{tab:cosmology_parameters} and~\ref{tab:nuisance_parameters}, and with $n(z)$ uncertainty marginalized using three different approaches. The green contour shows the results using the average $n(z)$ with no uncertainties, while the red and blue contours show the results using the direct and Bayesian resampling approaches  described in Section~\ref{sec:mtd:resample}. The orange contours use the shift model parameterization, with $N_{\rm tomo} = 4$ nuisance shift model parameters, described in Section~\ref{sec:mtd:shift}.  The dashed lines are the true values used to  generate data vector. This plot is made using \textsc{chainconsumer} \citep{Hinton2016} }
    \label{fig:full-2d}
\end{figure*}

\begin{figure}
    \centering
    \includegraphics[width=1.0\columnwidth]{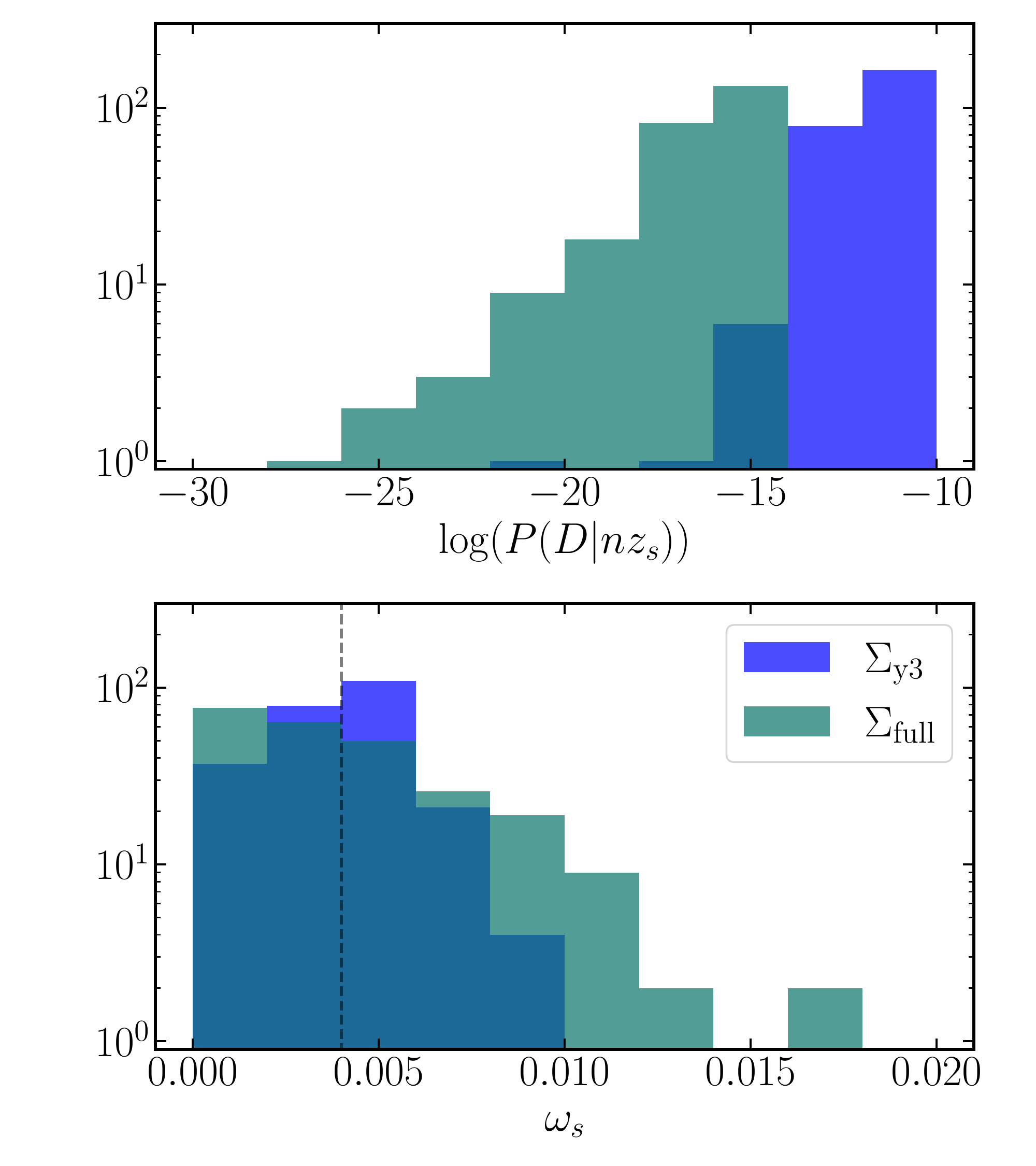}
    \caption{\textbf{Top row:} the distributions of the log-evidence $\log(P(\vect{D}|\nzs{,s}))$, defined in Eq.~\eqref{eq:bayesian_evidence}. \textbf{Bottom row:} the Bayesian weight, $\omega_s$, defined in Eq.~\eqref{eq:resampling_weights} and  applied to the chains in the Bayesian resampling approach. The vertical dashed line in the bottom panel is the constant weight applied to each chain in the `direct resampling' method, $\omega_c = 1/250$. The distributions of log-evidence and Bayesian weight are broader for  $\mt{\Sigma}_{\rm full}$ than for $\mt{\Sigma}_{\rm y3}$, because the same amount of $n(z)$ uncertainty has a larger impact on the more statistically powerful dataset, i.e., the full HSC dataset. }
    \label{fig:logz_weight}
\end{figure}

\subsubsection{Methods Summary}
\label{sec:mtd:summary}

In this section, we briefly summarize the methods for marginalizing over $n(z)$ uncertainty in this work, including the notation and terminology of the marginalization methods.

\begin{itemize}
    \item \text{No $n(z)$ Uncertainty}: We use the average histogram bin height of the $n(z)$ prior, \refresponse{$\langle \nz^i \rangle$}, as the sample redshift distribution, without marginalizing over any $n(z)$ uncertainties. This is the baseline that other methods are compared to. 
    \item \text{Direct Resampling}: We sample $N_{\rm sample}$ realizations of $\nzs{,s}$ from the $n(z)$ prior and run cosmological parameter inference on each realization without explicitly accounting for the evidence of the $\nzs{,s}$. The chains for different $\nzs{,s}$ are then combined with equal weights, implicitly incorporating the $n(z)$ uncertainties into the resulting parameter constraints. 
    \item \text{Bayesian Resampling}: This method begins as does direct resampling, but the chains for different $\nzs{,s}$ are weighted by their Bayesian evidence, as described in Section~\ref{sec:mtd:resample}.
    \item \text{Shift Model}: The average histogram bin heights $\langle \nz^i \rangle$ is allowed to shift on redshift individually for each tomographic bins, resulting in $N_{\rm tomo} = 4$ nuisance parameters for  marginalizing over redshift uncertainty, as described in Section~\ref{sec:mtd:shift}.
\end{itemize}

\subsection{Probability Integral Transformation}
\label{sec:mtd:pit}

In this section, we introduce our validation method for the parameter inference results. 
\refresponse{We note that validating the probability calibration of inference results is an integral part of testing novel inference methodology. Since the `true value’ of a parameter of interest is viewed in the Bayesian picture as a random variable, the posteriors derived using an inference methodology have to present an accurate estimate of that unknown distribution.}

\refresponse{A necessary requirement is that our inference adheres to Bayes theorem, which forms the basis of the statistical test presented in the following. To test this,} we perform a statistical test based on the probability integral transformation \citep[PIT;][]{casella2002statistical, 2020MNRAS.499.1587S} to test the validity of the inference statistics. We perform PIT on the cumulative density function (CDF) of $S_8$, as it is the parameter that the cosmic shear  constrains most precisely. The true posterior of the inferred $S_8$ can be yielded by Bayes' theorem:
\begin{equation}
    P(S_8|\vect{D}) = \frac{P(\vect{D}|S_8) P(S_8)}{P(D)}.
\end{equation}
We define the CDF of $S_8$ to be 
\begin{equation}
\label{eq:cdf_of_y}
F_{S_8}(S_8) = \int_0^{S_8} \mathrm{d} S_8' P(S_8'|\vect{D}).
\end{equation}
According to the PIT theorem, a random variable $Y$ drawn from the distribution of $F_{S_8}(S_8)$ in Eq.~\eqref{eq:cdf_of_y}, has a range of $[0,1]$, and the CDF of $Y$ follows
\begin{equation}
    F_Y(y) = y,
\end{equation}
where $y$ is a specific value of $Y$ between $[0,1]$.

To test the credibility of our inference pipeline, we estimate the CDF of $S_8$, namely, $\hat{F}_{S_8}(y)$, by generating pairs of data vectors $\vect{D}^\mu$ and $S_8^\mu$, where $\mu = 1,2\dots N_{\rm PIT}$, and $N_{\rm PIT} = 50$. For each $\mu$, we sample a pair of ($\Omega_m^{\mu}, \sigma_8^\mu$) with the uniform prior $U_{\Omega_m}[0.2,0.4]$ and $U_{\sigma_8}[0.7, 1.0]$, and compute the corresponding $S_8^\mu = \sigma_8^\mu \sqrt{\Omega_m^{\mu}/0.3}$. We first produce a noiseless data vector using the average $n(z)$, and then add a random noise realization generated using $\mt{\Sigma}_{\rm y3}$. The noisy data vector is denoted $\vect{D}^{\mu}$. 

We run the full inference pipeline on each pair of $\vect{D}^\mu$ and $S_8^\mu$, which generates a posterior $P^{\mu}(S_8|\vect{D}^\mu)$. For each $\mu$, we estimate
\begin{equation}
    \hat{Y}^{\mu} = F^\mu_{S_8}(S_8^\mu),
\end{equation}
where $F^\mu_{S_8}$ is the CDF of the $S_8$ posterior for the $\mu$-th sample. We compare the CDF of $\hat{Y}$ with the expected uniform distribution in Sec.~\ref{sec:res:pit}.

By conducting the PIT test, we are checking that the posterior distribution of the cosmological parameters inferred in the inference pipeline is statistically consistent with the true posterior given by Bayes' theorem. This is a crucial validation test for the results of this work, since our conclusion that compares marginalization methods relies on accurate posterior errorbars of the inferred parameters. Crucially, this test must be done using data vectors with noise added according to the covariance matrix, since that noise is what broadens the parameter distribution that we are trying to infer.


\section{Results}
\label{sec:res:0}


In this section, we show the results of forecasting cosmic shear analyses with different marginalization approaches, following the methods outlined in Sec.~\ref{sec:mtd:0}. In Section~\ref{sec:res:full-ana}, we show results of the full analyses, where 5 cosmological parameters, 2 IA parameters, 4 multiplicative biases, 2 PSF systematics parameters, and any parameters used to parametrize uncertainty in $n(z)$ are jointly fit. In Section~\ref{sec:res:pit}, we show the PIT validation on noisy data vectors. In Section~\ref{sec:res:comp}, we compare the results in this work to that of other work.

\subsection{Full analysis}
\label{sec:res:full-ana}


In this section, we show the results of the full cosmic shear analysis on the noiseless mock data vector using the redshift marginalization methods listed in Section~\ref{sec:mtd:summary}. We consider 5 cosmological parameters, listed in Table~\ref{tab:cosmology_parameters} and explained in Section~\ref{sec:mtd:cosmology}. Additionally, we consider 2 IA parameters, $N_{\rm tomo} = 4$ multiplicative biases, and 2 PSF systematics parameters, listed in Table~\ref{tab:nuisance_parameters} and explained in Section~\ref{sec:mtd:nuisance}. 

We ran a baseline analysis with the average $n(z)$ and no marginalization for comparison, and three marginalization approaches: the direct and Bayesian resampling, described in Section~\ref{sec:mtd:resample}, and the shift model, described in Section~\ref{sec:mtd:shift}. For the resampling approach, we ran $N_{\rm sample} = 250$ chains for both $\mt{\Sigma}_{y3}$ and $\mt{\Sigma}_{\text{full}}$ covariance matrices.  
There are $N_{\rm tomo} = 4$ nuisance parameters for the shift model, for which the fiducial values and priors are listed in Table~\ref{tab:shift_model_table}. 

In the top row of Fig.~\ref{fig:full-2d}, we show the 2-d posterior contours and their 1-d projections on the $\Omega_m$-$S_8$ plane for all four analyses, for the $\mt{\Sigma}_{y3}$ covariance (left), and $\mt{\Sigma}_{\text{full}}$ covariance (right). \refresponse{For the three-year HSC analyses, the different methods of redshift marginalization do not make a visible difference in the contour plot.} However, the contours are visibly different for the future full data set of HSC. For $\mt{\Sigma}_{y3}$, the number of resampling for both covariances are $N_{\rm sample,y3} = N_\text{\rm sample,full} = 250$. 

In Fig.~\ref{fig:logz_weight}, we show the distribution of log-evidence $\log(P(\vect{D}|\nzs{,s}))$ and the Bayesian weight $\omega_s$, defined in Eq.~\eqref{eq:bayesian_evidence} and Eq.~\eqref{eq:resampling_weights}, of the chains in the resampling approach. The direct resampling method  applies uniform weights, while the Bayesian resampling method  applies the Bayesian weights $\omega_s$. Since the HSC full data-set has a three-times smaller covariance matrix than the three-year data-set, the same $n(z)$ uncertainty causes a more significant scatter in both the log-evidence and Bayesian weight. This means that Bayesian resampling will become increasingly favoured over direct resampling as the dataset becomes more statistically powerful.  
In practice, the Bayesian resampling approach is assigning more weight to realizations of the $n(z)$ that produce data vectors that are more consistent with the expected one, while down-weighting realizations with less evident $n(z)$.

\begin{figure*}
    \centering
    \includegraphics[width=1.0\columnwidth]{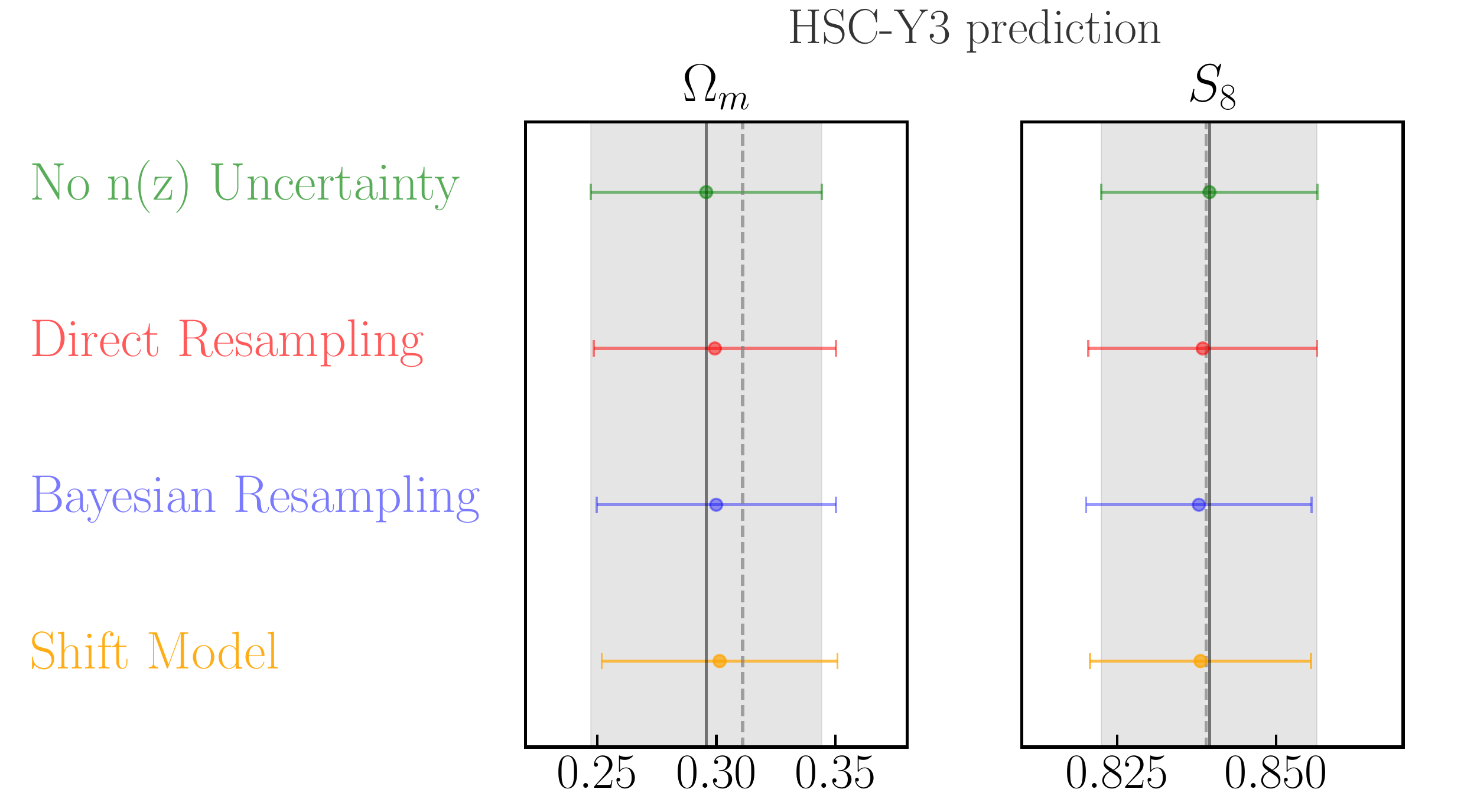}
    \includegraphics[width=1.0\columnwidth]{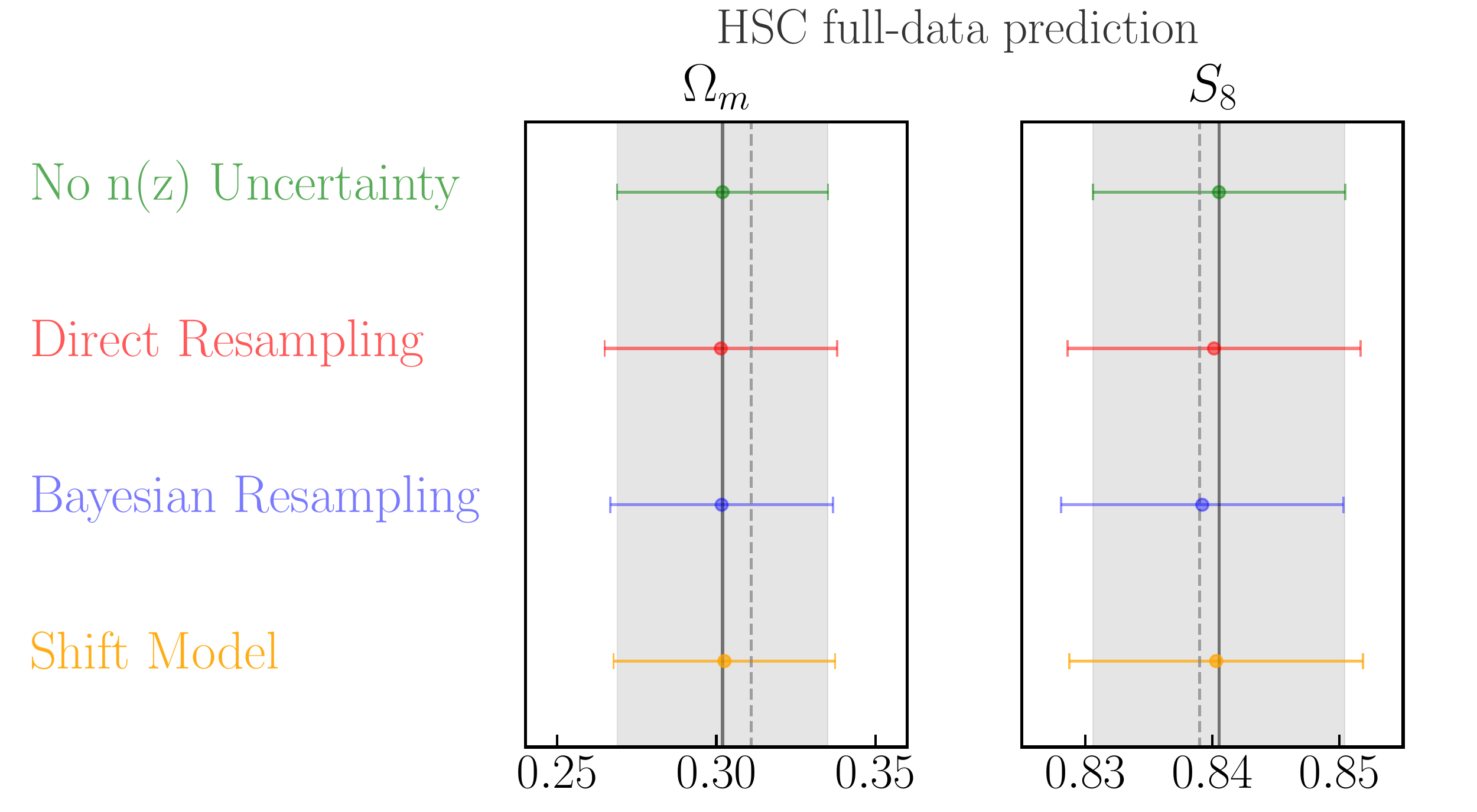}
    \caption{ The $68\%$ confidence intervals (bars) and mean values (dots) of the 1-d projection for the $n(z)$ marginalization approaches compared to the `No $n(z)$ uncertainty' run, for the $\mt{\Sigma}_{y3}$ covariance matrix (left), and $\mt{\Sigma}_{\text{full}}$ covariance matrix (right), for the full analysis. The solid reference lines and the shaded area are the mean values and the $68\%$ confidence intervals of the ``No $n(z)$ uncertainty'' run. The dashed lines are the parameter truth in Table~\ref{tab:cosmology_parameters}. We find that the mean values of $\Omega_m$ are systematically lower than the true input across different methods. The could be caused by the skewness of the posterior distribution.  
    }
    \label{fig:full-1d}
\end{figure*}

\begin{figure}
    \centering
    \includegraphics[width=1\columnwidth]{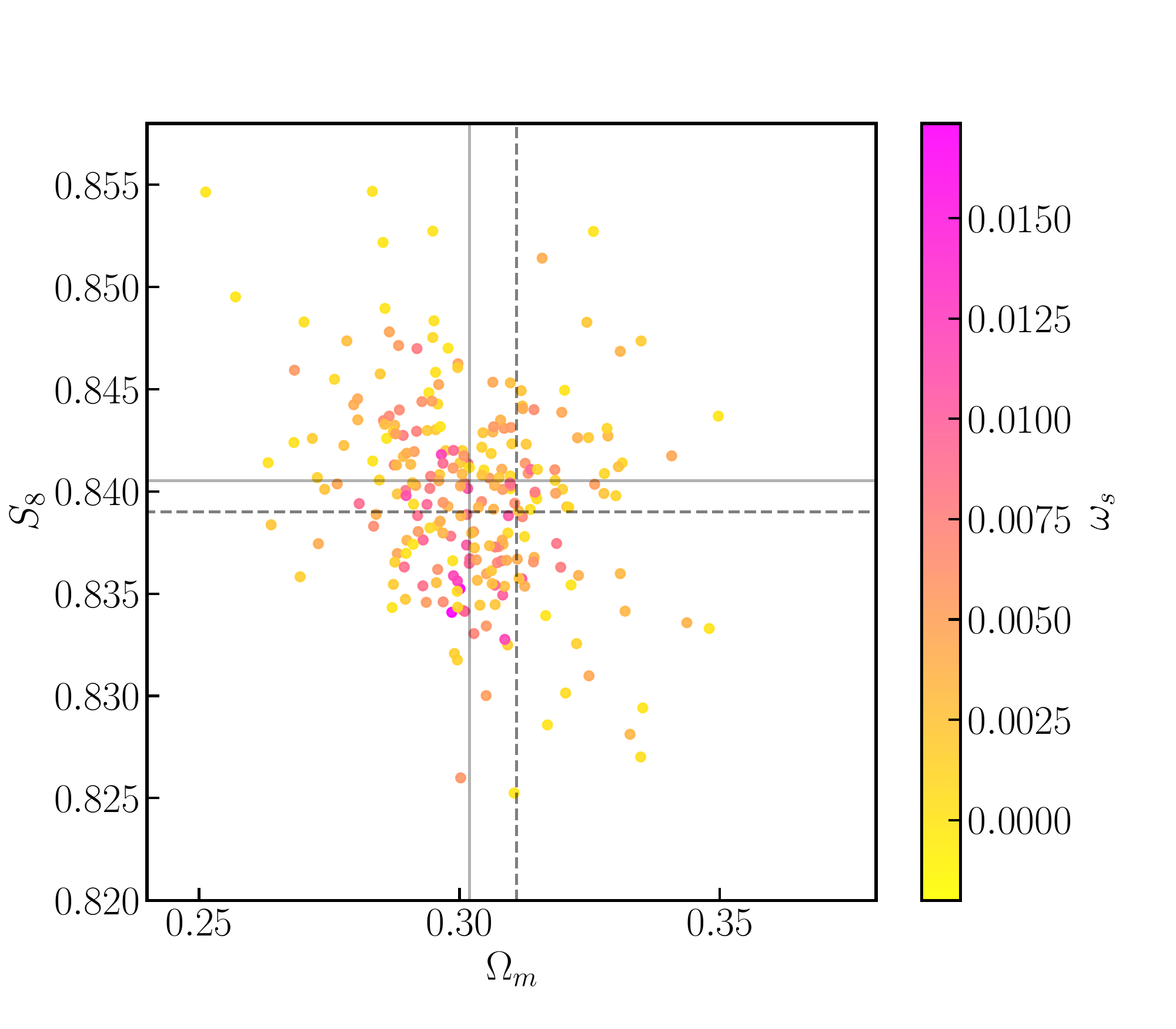}
    \caption{A scatter plot showing  the mean posterior values for the $N_{\rm sample} = 250$ chains 
    with different samples drawn from the $n(z)$ prior, analyzed with all 17 parameters (using the shift model) and with  $\mt{\Sigma}_{\text{full}}$.  The colors indicate the Bayesian weight $\omega_s$ that the chain receives, determined by its evidence $P(\vect{D}|\nz)$, defined in Eq.~\eqref{eq:bayesian_evidence}. The solid line represents the mean statistics of the `No $n(z)$ uncertainties' run, while the dashed line represents the true input parameters.   
    }
    \label{fig:200chains}
\end{figure}

In Fig.~\ref{fig:full-1d}, we show the  uncertainty for individual cosmological parameters from the full analysis chains in Fig.~\ref{fig:full-2d}. We used the mean parameter value as the point estimation and the $68\%$ confidence interval as the error bars of the ``No $n(z)$ uncertainty" run for the reference. We also show the true value of the parameters in dashed line, as a comparison. For the three-year analyses, shown on the left, marginalizing over the redshift distribution uncertainty does not noticably increase the error bars on either $\Omega_m$ and $S_8$, except when using the ``Direct resampling" method. 
Since the Bayesian resampling method provides a principled approach to incorporation of redshift distribution uncertainties, we take the consistency between that method and the no marginalization method as a sign that the uncertainty in the cosmic shear data vector dominates the uncertainties on cosmological parameters.  Therefore, the ``Direct resampling" may be introducing spurious uncertainty by failing to down-weight $n(z)$ realizations that are inconsistent with the data vectors, and is not recommended. For the full HSC dataset analyses, shown on the right, we can see that the conclusion of the three-year analyses holds, though the differences between the methods are more visible. The mean posteriors of the $\Omega_m$ are systematically lower than the true input value across different methods. We suspect that the banana-shaped $\Omega_m - \sigma_8$ degeneracy that occurs in the full analysis skews the projected distribution of $\Omega_m$ to the lower end, which also causes the underestimation of $\Omega_m$ in Fig.~\ref{fig:200chains}.

\begin{figure}
    \centering
    \includegraphics[width=1.0\columnwidth]{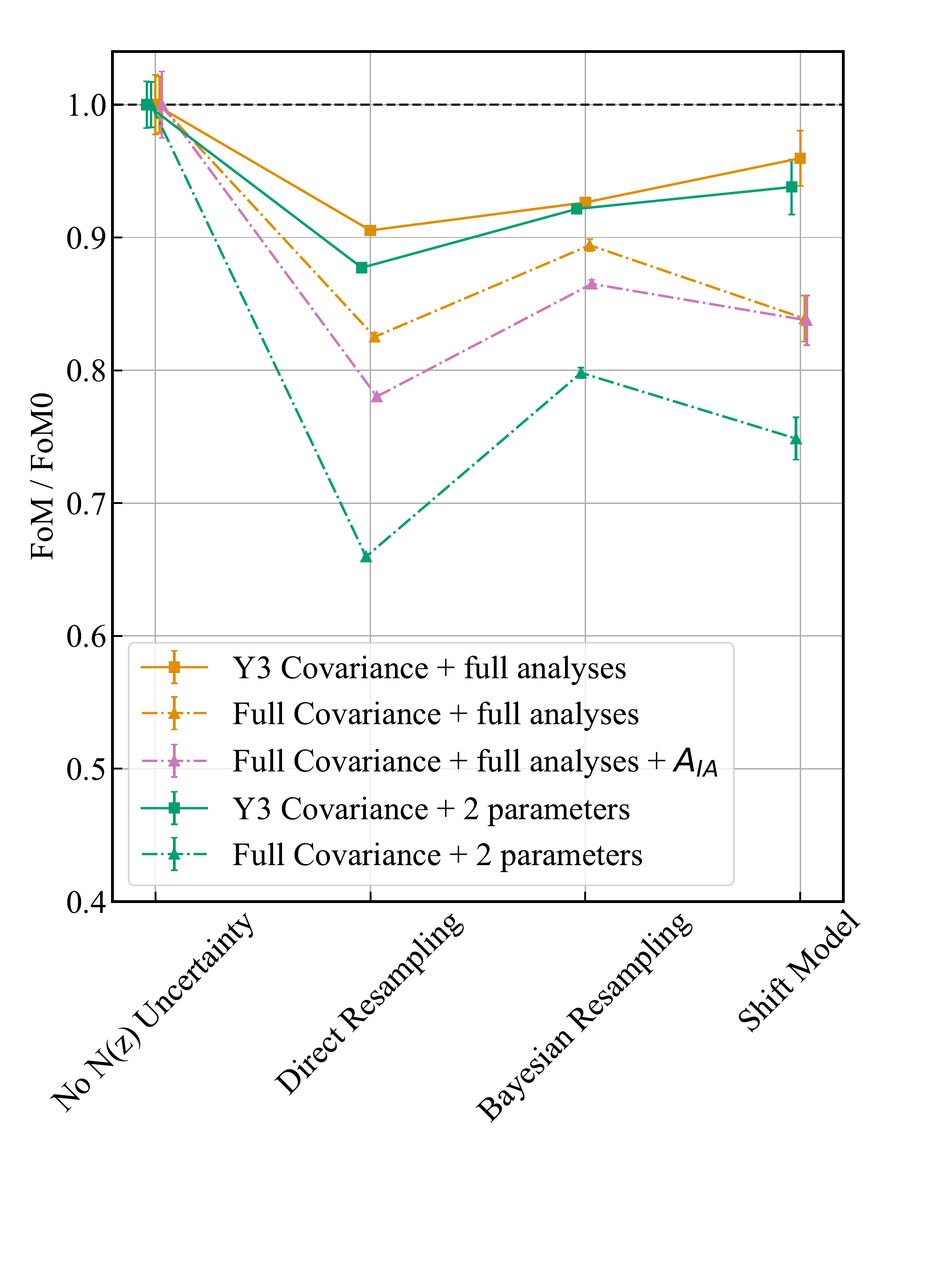}
    \caption{ The Figure of Merit (FoM) comparison in the $\Omega_m$-$S_8$ plane for each marginalization method. The uncertainties on the FoM are computed by bootstrapping the MCMC chains 100 times. All FoMs are divided by the FoM of the ``No $n(z)$ uncertainty" run. \refresponse{The purple line shows FoM ratio for the 3D $\Omega_m$-$S_8$-$A_{\rm IA}$ space, while other lines shows the FoM in the $\Omega_m$-$S_8$ space.} The direct resampling method is clearly the most conservative method of those tested in this work. The shift model shows similar performance to the Bayesian resampling method. \refresponse{The errorbars on the FoM are given by bootstrapping the chains, which matches the errorbars given by running the analysis with different sampling seeds.}}
    \label{fig:all_fom}
\end{figure}

We further computed the Figure of Merit (FoM) in the $\Omega_m$-$S_8$ plane (or $\Omega_m$-$S_8$-$A_{\rm IA}$ space) to compare the methods, defining the FoM as 
\begin{equation}
    \text{FoM} = \frac{1}{\sqrt{|\mt{F}^{-1}|}},
\end{equation}
where $\mt{F}$ is the Fisher matrix of $[\Omega_m, S_8]$(or $[\Omega_m, S_8, A_{\rm IA}]$). $\mt{F}$ is calculated by taking the inverse of the covariance matrix of $[\Omega_m, S_8]$ (or $[\Omega_m, S_8, A_{\rm IA}]$), approximating the \textsc{MultiNest} posterior as a 2(3)-d Gaussian distribution. This approximation effectively marginalizes over the other parameters that are varied during the parameter inference. 
The FoM is proportional to the reciprocal of the contour area. In Fig.~\ref{fig:all_fom}, we plot the FoM of all the marginalization methods, divided by the FoM value of the ``No $n(z)$ Uncertainty". The two oranges lines correspond to the full analyses in this section.
Unsurprisingly, the direct resampling method provides more conservative parameter constraints compared to the Bayesian resampling method, since it does not downweight the outlier $n(z)$ realizations even though they are unlikely to produce the observed shear data vector. The shift model is slightly conservative for $\mt{\Sigma}_\text{full}$, and slightly optimistic for $\mt{\Sigma}_{y3}$, compared to the Bayesian resampling. \refresponse{The errorbars on the FoM values are obtained by bootstrapping the chains. As a cross-check on our errorbars, we also ran 10 chains using the shift model for the Y3 analysis, with different sampling seeds. The errorbars obtained using the standard deviations of the inferred cosmological parameters using these 10 chains is within 5\% of those from bootstrapping, which suggests that seeding noise cannot explain the differences in FoM between the methods. }

\refresponse{Additionally, we report the ratio of FoMs to the fiducial one in the 3D $\Omega_m$-$S_8$-$A_{\rm IA}$ space using the full covariance matrix $\mt{\Sigma}_{\rm full}$. Since the amplitude of intrinsic alignment is also sensitive to the redshift distribution, different marginalization methods also impact its constraints. The FoM in the $\Omega_m$-$S_8$-$A_{\rm IA}$ space (purple line) follows the same trend as the orange dashed lines in Figure~\ref{fig:all_fom}, however, the difference between Bayesian resampling and shift model decreases from $3\%$ to $1\%$ of 1-$\sigma$, while the difference between the Bayesian resampling and direct resampling decreases from $4\%$ to $3.3\%$ of 1-$\sigma$\footnote{$\Delta$FoM/FoM0 is proportional to $-2\Delta \sigma/\sigma_0$ for two parameters, while $\Delta$FoM/FoM0 is proportional to $-3\Delta \sigma/\sigma_0$ for three parameters, where $\sigma_0$ is the $<68\%$ confidence range of `no marginalization'}. This further strengthens the conclusion that the Bayesian resampling method behaves comparably to the shift model in HSC Y3 cosmic shear analyses, while direct resampling tends to overestimate the uncertainty in the parameter constraints. }

\refresponse{Finally, Fig.~\ref{fig:all_fom} also shows a FoM comparison for an analysis with only two free cosmological parameters, $\sigma_8$ and $\Omega_m$, rather than with all cosmological parameters free.  For more details of this analysis, see Appendix~\ref{app:res:2p-exp}.  For this more limited analysis, the direct resampling method overestimates the uncertainties in the $(\Omega_m, S_8)$ plane compared to the Bayesian resampling, and therefore is not recommended. The shift model is slightly conservative in this more limited analysis for the full dataset, and slightly optimistic for the three year analysis. }

In Fig.~\ref{fig:200chains}, we show the 1-d mean posterior points of 250 chains in the resampling approach, run with $\mt{\Sigma}_{\text{full}}$. 
The color of the points are coded by the Bayesian weight $\omega_s$ of the chain, which is proportional to the model evidence $P(\vect{D}|\nz)$. We can see that drawing different samples from the $n(z)$ posterior introduces scatter in the mean values in the $\Omega_m$-$S_8$ plane, but generally the samples with mean closer to the centre of the cluster receive a higher weight, while the $n(z)$ samples that generate outliers are down-weighted. This plot demonstrates the necessity of considering whether a given $n(z)$ sample is likely to have generated the data vector that we are observing during the resampling process -- as is done in the Bayesian resampling approach, but not direct resampling. We notice that there are $nz_s$ samples that generate mean posterior at the centre of the cluster, but receive a very low weight. There are two possible explanations: (a) the realization $nz_s$ has a best-fit data vector that is on average unbiased compared to the mock data vector $\vect{D}$, but for certain redshifts or $\theta$ values there are significant deviations (with opposite signs, so they compensate on average); (b) the best-fit data vector deviated from that for the true cosmological parameters in a way that is compensated by biases in other cosmological parameters besides $\Omega_m$ and $S_8$.  
The mean values of the $\Omega_m$ are systematically lower than the true value of the input, as we explained earlier in this section.

\begin{table*}
\begin{tabular}{lllllll}
\hline
Method                & Live Points & Efficiency & \refresponse{Tolerance} & \# of chains & CPU-hour/chain & total CPU-hour \\\hline
No $n(z)$ uncertainty & 500         & 0.1    & \refresponse{0.2}    & 1           & 1.76h*56       & 98.9h          \\
Direct Resampling     & 200         & 0.1    & \refresponse{0.2}    & 250         & 1.05h*28       & 7350.0h          \\
Bayesian Resampling   & 200         & 0.1    & \refresponse{0.2}    & 250         & 1.05h*28       & 7350.0h          \\
Shift Model           & 500         & 0.1    & \refresponse{0.2}    & 1           & 1.77h*56       & 99.1h         \\\hline
\end{tabular}
\caption{\refresponse{The  \textsc{MultiNest} settings used in this work}, and computational expense for different marginalization methods, for the full analysis using $\bm{\Sigma_{y3}}$. The chains are run on Vera, a dedicated server for the McWilliams Center for Cosmology. 
Each node is equipped with 2 Intel Haswell (E5-2695 v3) CPUs, which have 14 cores per CPU. The resampling approaches, due to the need to run hundreds of individual analyses, are two orders of magnitude slower than the shift model. \refresponse{All chains are ran in constant efficiency mode for more accurate evidence estimation.}
}
\label{tab:time_comparison}
\end{table*}

Following the above presentation of the analysis results, we also compare the computational performance of each redshift distribution marginalization method. In Table~\ref{tab:time_comparison}, we show the  \textsc{MultiNest} settings used for each  method, and the computational expense of the full analysis in CPU-hours. The resampling approaches are two orders of magnitude slower than the shift model. 
While the Bayesian resampling and shift methods lead to comparable uncertainties, as is shown in Fig.~\ref{fig:all_fom}, the tremendous computational efficiency of the shift model compared to the Bayesian resampling makes it the recommended choice for the HSC three-year analyses.

For the full HSC three-year cosmic shear analysis, our results suggest that the shift model will produce uncertainties on cosmological parameters that are consistent with the principled Bayesian resampling method to within $3\%$ \refresponse{of 1-$\sigma$}. Considering that the orders of magnitude difference in computational expense, we recommend the shift model as a well-understood and sufficiently accurate approach for the HSC three-year analysis.

\subsection{Inference Validation}
\label{sec:res:pit}

\begin{figure}
    \centering
    \includegraphics[width=1\columnwidth]{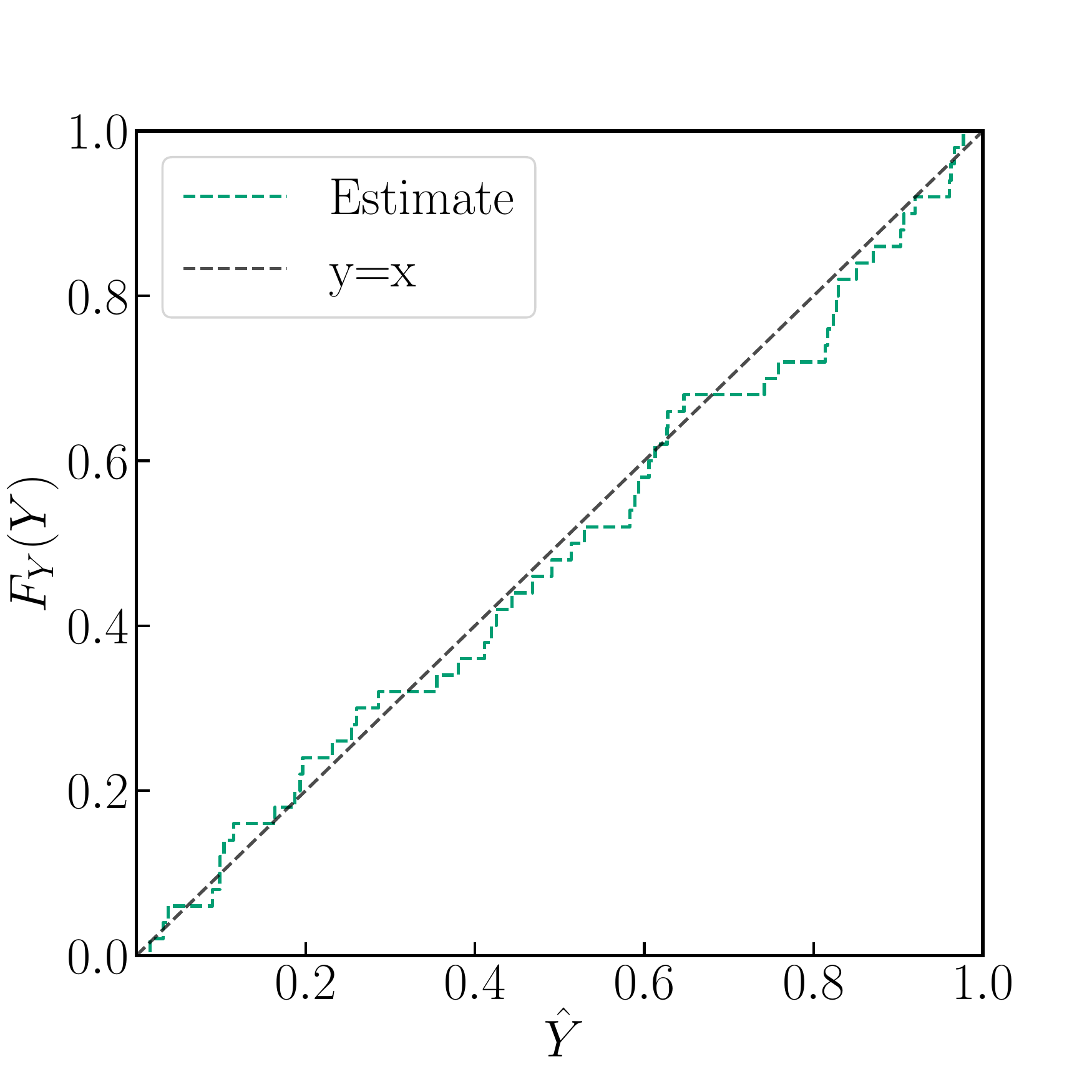}
    \caption{The CDF of $\hat{Y}$, defined in Sec.~\ref{sec:mtd:pit}. If the statistical inference preserves Bayes' theorem, the CDF of $\hat{Y}$ should follow $y=x$, which is plotted as the black dashed line. This plot shows a high level of consistency between the estimated and  expected distribution of $\hat{Y}$. 
    }
    \label{fig:pit_saw}
\end{figure}

In this section, we present the inference validation by performing the probability integral transformation (PIT), as described in Sec.~\ref{sec:mtd:pit}. 
\refresponse{We will focus our analysis on the shift model since it represents the simplest methodology that is appropriate for our data as described in the previous sections. While computationally more expensive, we could also perform the same test for the Bayesian and direct Resampling methods. Given that the three aforementioned methods perform similarly in the context of HSC Y3 analysis, we defer a more detailed investigation to future work and concentrate here on the shift model case. }

We sample $N_{\rm pit} = 50$ $\Omega_m$-$\sigma_8$ pairs, generate a corresponding noisy data vector, obtain the marginalized $S_8$ posterior from the full inference with shift model, and compute the CDF of the corresponding true $S_8$ values. 

In Fig.~\ref{fig:pit_saw}, we compare the CDF of $\hat{Y}$, the CDF of $S_8$ evaluated at the true $S_8$, 
with the CDF of an expected uniform distribution, shown in the black dashed line. On visual inspection, the estimated CDF follows the expected $y=x$ line nicely. We also conduct an Kolmogorov–Smirnov (K-S) test, which computes the maximum difference between the CDF and the expected CDF. The K-S results is $D=0.094$, with a $p$-value of $ 0.737$. This means that $\hat{Y}$ is highly consistent with the uniform distribution, which validates our inference pipeline.

\subsection{Literature Comparison}
\label{sec:res:comp}

In this section, we compare our Y3 results with the results of marginalizing over $n(z)$ uncertainties in other cosmic shear analysis works.  This comparison necessarily excludes the Bayesian resampling approach outlined in this paper, as to the authors' knowledge it has not previously been applied.

In \cite{2017MNRAS.465.1454H}, direct resampling marginalization is tested using 3 different $n(z)$ uncertainty distributions: weighted direct calibration \citep[DIR;][]{Lima_2008}, angular cross-correlation calibration \citep[CC;][]{2008ApJ...684...88N}, and a recalibration of $p(z)$ estimated by \textsc{bpz} \citep[BOR;][]{2010MNRAS.406..881B}.  
Compared to `no $n(z)$ marginalization', the DIR, CC, and BOR approaches increase the uncertainty on $S_8$ by $14\%, 90\%$, and $19\%$ \refresponse{of 1-$\sigma$}, respectively. In comparison, we find that direct resampling increases the $S_8$ errorbar by $5.8\%$ \refresponse{of 1-$\sigma$} for $\mt{\Sigma}_{\rm y3}$ , and $15.3\%$ \refresponse{of 1-$\sigma$} for $\mt{\Sigma}_{\rm full}$. This smaller increase in uncertainty is likely due to the larger covariance matrix and a tighter $n(z)$ prior for the HSC Y3 analysis. 

In \cite{2020PASJ...72...16H}, the shift model is adopted as the fiducial approach to marginalizing over $n(z)$ uncertainty, and is compared with `no $n(z)$ marginalization'. The uncertainty on $S_8$ increased by $1.4\%$ \refresponse{of 1-$\sigma$} after marginalizing over the $n(z)$ uncertainties with the shift model, with a wider prior than the one in this work. In this work, the errorbar on $S_8$ increased by $2.0\%$ \refresponse{of 1-$\sigma$} after marginalizing using the shift model. Given that \cite{2020PASJ...72...16H} has a larger covariance on the shear data vector, as well as a larger prior on the shift model parameters, the results should not agree exactly, and there is no reason to believe they are inconsistent. 

In \cite{2018PhRvD..98d3528T}, the shift model is compared with `no $n(z)$ marginalization'. The prior on the shift model parameters are comparable to this work, while the covariance matrix of \cite{2018PhRvD..98d3528T} is smaller than this work.  \cite{2018PhRvD..98d3528T} found a $4.4\%$ \refresponse{of 1-$\sigma$} increase in the $S_8$ uncertainty, which slightly larger than this work.  Given the differences in the data covariance for the analyses, the fact that marginalization had a greater impact in \cite{2018PhRvD..98d3528T} is consistent with expectations. 

In \cite{2022PhRvD.105b3514A}, the shift model is compared with a more sophisticated $n(z)$ marginalization method, called \textsc{hyperrank} \citep{2022MNRAS.511.2170C}. The shift model is found to be sufficient for cosmic shear analyses for DES Y3, 
as validated by \textsc{hyperrank}. The fact that a current survey found the shift model to be sufficient  is consistent with our finding for HSC Y3. 

In \cite{2021A&A...650A.148S}, a self-calibrated method that models the histogram bin heights $\nz$ as a series of comb Gaussian functions is used to analytically marginalize over the $n(z)$ uncertainties. The results are compared to the analysis in \cite{2020A&A...640L..14W}, which uses a shift model. There are only $1\%$ differences in $\chi^2$ between the results from the self-calibration method and the shift model, though there is a $10\%$ \refresponse{of 1-$\sigma$} difference in the intrinsic alignment amplitude $A_{IA}$. This once again shows that the shift model is sufficient for the current generation of cosmic shear analysis  for the purpose of cosmological parameter inference, which is consistent with our conclusion.


\subsection{Summary of results}
\label{sec:res:sum}

Overall, our results show that the shift model is a computationally efficient and credible marginalization method for the HSC three-year analysis. Therefore, we recommend that the HSC three-year analysis adopt the shift model for marginalizing $n(z)$ uncertainty. 

For the resampling approaches, we find that using the direct resampling approach consistently results in larger contours compared to the Bayesian resampling, as expected. Therefore, we suggest future cosmological analysis adopt the Bayesian resampling method, if resampling is necessary. 

For cosmic shear analyses with a substantial uncertainties on the sample redshift distribution, we recommend comparing any candidate marginalization methods for $n(z)$ with the results from the Bayesian resampling method, as the Bayesian resampling method provide a statistically-principled posterior on the marginalized parameters.

\section{Conclusions}
\label{sec:conclu}

The goal of this work was to understand the performance of methods for incorporating uncertainty in the ensemble redshift distribution in cosmological weak lensing shear analyses, including their impact on computational expense and on the estimated uncertainties on cosmological parameters.

We proposed a statistically-principled method, called Bayesian resampling, for marginalizing over the uncertainties of the sample redshift distribution $n(z)$ in the cosmic shear analysis. By adding a weight proportional to the model evidence of each $n(z)$ realization, Bayesian resampling effectively down-weights those realizations that are unlikely to generate the observed cosmic shear data vector. The Bayesian resampling method can be applied to any $n(z)$ uncertainties that can be modeled by a probability distribution, even if such parameterization is at a high dimensionality that makes it impossible to model in MCMC. 

We ran mock analyses for the HSC three-year and full-data cosmic shear, with 3 $n(z)$ marginalization methods: (a) the newly developed Bayesian resampling method; (b) the direct resampling, for which the weights of all $n(z)$ realizations are the same; (c) the shift model, the most prevalent parameterization used in cosmic shear analyses. Additionally, we ran analyses without marginalizing over the $n(z)$ as a comparison. Our mock data vector is the average cosmic shear signal from the fiducial cosmology, and its covariance is estimated by reducing the covariance compared to that in \cite{2020PASJ...72...16H} 
to account for survey area increases, for the three-year analysis and full analysis correspondingly. Our full theoretical model consists 5 $\Lambda$CDM parameters, 2 intrinsic alignment parameters, 4 multiplicative biases and 2 PSF systematics parameters, plus the 4 redshift parameters when the shift model is adopted. 

We compared the 3 marginalization methods and the analysis without marginalization in terms of their impact on the $\Omega_m$-$S_8$ contours, their 1-d errorbars, the figure of merit (FoM), and computational cost. Here is a high-level summary of how the methods compared to each other.
\begin{itemize}
    \item Marginalizing over $n(z)$ uncertainties yields larger errorbars on both $\Omega_m$ and $S_8$ for all methods. 
    \item Bayesian resampling yields significant tighter errorbars than direct resampling, implying that the direct resampling is overly-conservative for marginalization. 
    \item The shift model produces consistent errorbars to the Bayesian resampling for HSC Y3. Given that the computational cost for the shift model is $\sim 100$ times less, it is the recommended method for the upcoming HSC Y3 cosmic shear analyses. For the HSC full analysis, the shift model can yield errorbars that differ by $\sim 3\%$ \refresponse{of 1-$\sigma$} compared to Bayesian resampling, so it is unclear even in that case  whether alternative methods are worthwhile.
    \item \refresponse{Although the differences between the marginalization methods are statistically evident, the visual differences in the parameter constraint contours are not particularly noticeable. }
\end{itemize}

To test the credibility of our inferred  posterior probability distributions of cosmological parameters, we conducted the probability interval transformation (PIT) test on noisy data vectors generated with a range of cosmological parameters, to ensure the applicability of our results to real cosmic shear analyses. We sampled 50 pairs of $\Omega_m$-$\sigma_8$-$\vect{D}$, and compare the CDF distribution of $S_8$ at the true $S_8^\mu$ values with a uniform distribution. Our estimated CDF distribution passes the K-S test, thus validating our inference pipeline using the shift model. 

These results have a few implications for  future cosmic shear analyses. First, our results suggest that the shift model should be compared with Bayesian resampling for specific survey scenarios (statistical constraining power, etc.) to assess whether the shift model performs sufficiently well to be usable, given its far lower computational expense. The shift model is fundamentally a different $n(z)$ uncertainty model from the original $n(z)$ distribution.  Second, when using the resampling approach to marginalizing over $n(z)$ uncertainties is necessary for a weak lensing measurement, the Bayesian resampling approach is preferred over direct resampling, because of its consistency with  Bayesian statistics. 
Moreover, Bayesian resampling does require an accurate estimate of the ratio of the Bayesian evidences between realizations of redshift distributions.


There are several caveats in this work. (a) We reach the conclusion that a sophisticated marginalization method is going to be increasingly preferred based on the assumption that lensing measurements become more powerful as survey area increases, but the uncertainty on $n(z)$ is presumed to be systematics dominated.\refresponse{ The reason for this assumption is that the $n(z)$ uncertainties are limited by the cosmic variance of the COSMOS2015 field, which we used to assess the modeling uncertainties. If this assumption changes, then the comparison needs to be revisited. This assumption is discussed in detail in Section~\ref{sec:bgd:nz_estimation}. }  (b) We use the same angular and tomographic binning for the mock analyses in this paper, though the actual analyses of HSC Y3 and full data are likely to have different binning strategies. We also make very simple estimates of the covariance matrices in the mock analyses, ignoring the evolving footprint shape of the HSC survey. (c) The assumption in this work is that we can place a prior on the source redshift distribution that is statistically independent of our data vector. That was a good approximation  in this case, for $n(z)$ calibration based on photometry and cross-correlations, and for the data vector involving shear-shear only.  However, future analyses with more complex data vectors (e.g., including large-scale structure clustering) and/or $n(z)$ posteriors may violate this assumption in our formalism, which would require additional efforts to take into account. 

We conclude by mentioning some avenues for future investigations.  First, the cosmic shear data vector is sensitive to the mean redshift of the tomographic bin, which is likely the reason why the shift model is  sufficient for current surveys in practice. However, galaxy clustering is sensitive to other statistics of the ensemble redshift distribution, such as its width \citep[e.g.,][]{2022PhRvD.105b3520A}.  
Therefore, the validity of the shift model in  galaxy-galaxy lensing, clustering and 3x2pt analyses should be directly tested.

Finally, the resampling approach for the $n(z)$ marginalization requires thousands of CPU-hours. Importance sampling methods can be added to the method to reduce the number of realizations needed. However, importance sampling faces other challenges: since $n(z)$ distributions normally are parameterized with  high dimensionality, the importance weights are easily dominated by a few samples. It might also be extremely challenging to perform importance sampling on some $n(z)$ priors. It would be valuable to identify solutions to this problem and demonstrate how to effectively accelerate $n(z)$ resampling approaches using importance sampling. 

\section*{Contributors}

TZ developed the mock analysis pipeline, conducted the cosmology inferences and validation tests, and led the writing of the manuscript. MMR proposed the project, advised on the experimental design and analysis, provided and wrote about the sample redshift distribution data, and provided feedbacks on the results. RM advised on the motivation, experimental design and analysis, and edited the manuscript. XL provided implementation guidance on the inference pipeline and feedbacks to the results. BM provided comments and feedback to the results, and edited the manuscript. 

\subsection*{Acknowledgments}

\refresponse{We thank the anonymous referee for constructive feedback on this work. We thank Alex Malz, Chad Schafer, Andresa Campos, Biwei Dai, Danielle Leonard for their helpful discussion with us. }

TZ and RM are supported in part by the Department of Energy grant DE-SC0010118 and in part by a grant from the Simons Foundation (Simons Investigator in Astrophysics, Award ID 620789). 

We thank the developers of \textsc{CosmoSIS}, \textsc{NumPy}, and \textsc{ChainConsumer} for making their software openly accessible. 


\section*{Data Availability}

The data vector, covariance matrix and stacked $p(z)$ distribution are available on \url{http://th.nao.ac.jp/MEMBER/hamanatk/HSC16aCSTPCFbugfix/index.html}.
The sample redshift distribution, the cosmology inference and analysis code will be shared on reasonable request to the authors.


\bibliography{main}

\appendix

\section{Impact of $\xi_\pm(\langle \phi_{\rm nz} \rangle)$}
\label{sec:ap:nzave}

\begin{figure}
    \centering
    \includegraphics[width=1\columnwidth]{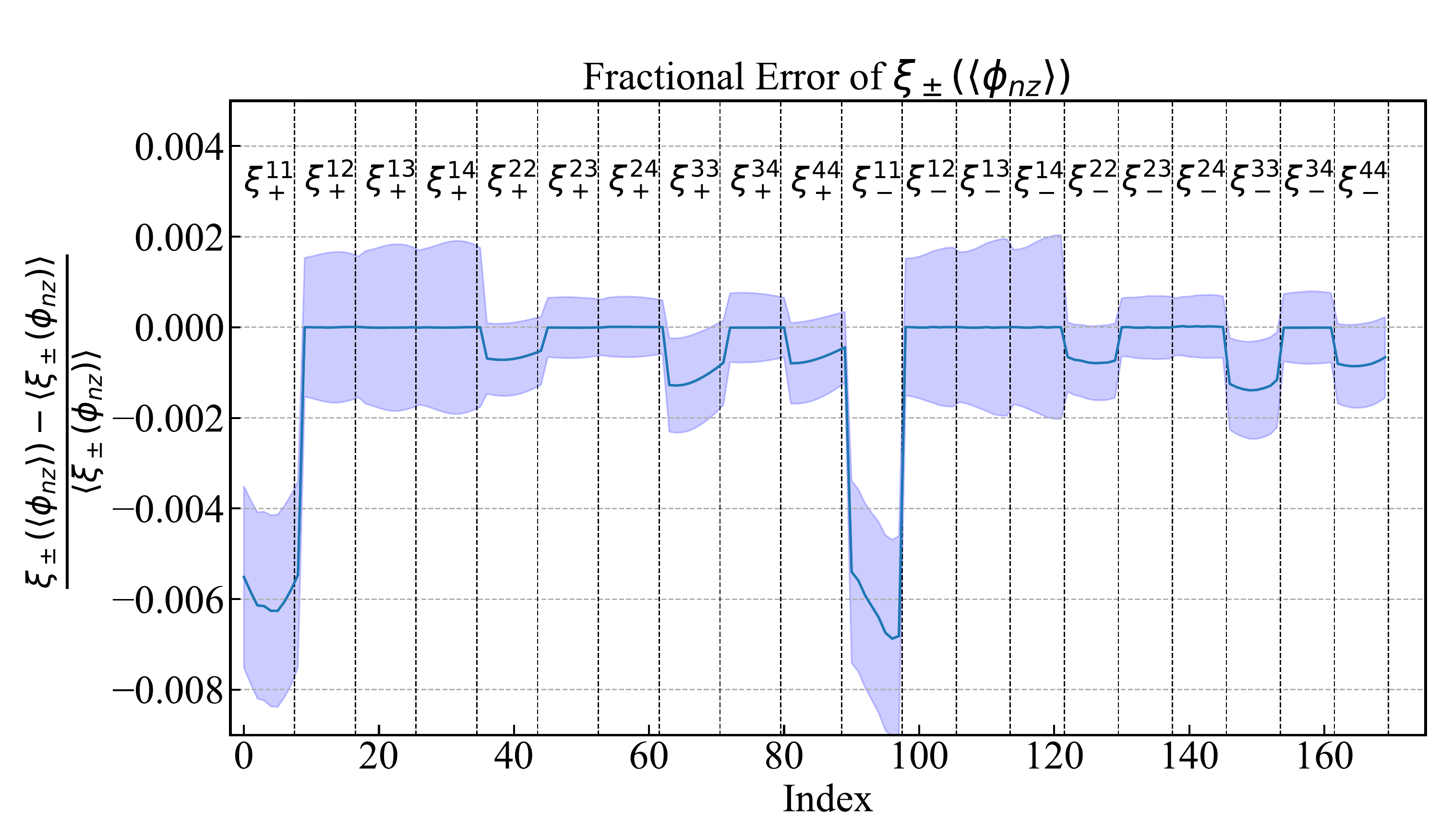}
    \caption{ \refresponse{The fractional error of the $\xi^{ij}_\pm(\langle n(z) \rangle )$ compared to $\langle \xi^{ij}_\pm(n(z) ) \rangle$. The x-axis is the index of the value in the data vector, with 9 angular bins for each tomographic $\xi_+^{ij}$ bin pair, and 8 angular bins for each $\xi_-^{ij}$ bin pair.  There is a statistically significant sub-percent difference between the auto-correlations, which shows that taking the average of $n(z)$ does not commute with computing the 2PCF.  Therefore $\langle \xi^{ij}_\pm(n(z) ) \rangle$ should be used to avoid sub-percent bias when this level of precision matters.  This fractional difference is largest at low redshift because the uncertainties in the mean redshifts are comparable in magnitude for all bins, but the 2PCF is lower at low redshift.} }
    \label{fig:frac_error_average}
\end{figure}

\refresponse{
In Figure~\ref{fig:frac_error_average}, we demonstrate that generating the auto-correlation of the mock data vector using the average $n(z)$ is different from the taking the average of data vectors generated by random draws from the posterior for $n(z)$. Therefore, for this work, in which the conclusion is sensitive to bias at the sub-percentage level, we choose to use $\langle \xi^{ii}_\pm(n(z) ) \rangle$ from 1000 $n(z)$ samples to generate the mock data vector in Section~\ref{sec:mtd:mock}.} 

\refresponse{
The reason that only the auto-correlation is affected in Figure~\ref{fig:frac_error_average} can be explained by Eqs.~\eqref{eq:angular_power} and~\eqref{eq:transfer_function}. Since $n^i(\chi(z))$ is independent of $n^j(\chi(z))$ if $i\neq j$, the transfer functions $q^i(\chi)$ and $q^j(\chi)$ are thus independent. As a result, when $i \neq j$, 
\begin{equation}
    \label{eq:angular_power_ave}
    \langle C_\ell^{ij} \rangle = \int \frac{\mathrm{d}\chi}{\chi^2} P(\ell/\chi;z(\chi)) \langle q^i(\chi)\rangle \langle q^j(\chi) \rangle. 
\end{equation}
Notice that Eq.~\eqref{eq:angular_power_ave} only holds when $n^i(\chi(z))$ is independent of $n^j(\chi(z))$. Otherwise, both auto- and cross-correlations in the mock data vector will be affected by using the average $n(z)$.  Also note that in the case that some overall source of uncertainty were to lead to correlations between the uncertainties in the redshift distributions for different bins, both auto- and cross-correlations would be affected.}

\refresponse{
\section{Two-Parameter Analyses}
\label{app:res:2p-exp}}

\refresponse{In this work, we also carried out the cosmological parameter inference for a case where only $\Omega_m$ and $\sigma_8$, along with $n(z)$ marginalization nuisance parameters, are freed. This scaled-down test is initially designed for testing and sanity-checking our inference and analysis software.  
The constant values for other cosmological and nuisance parameters, and the priors for $\Omega_m$ and $\sigma_8$, are listed in Tables~\ref{tab:cosmology_parameters} and~\ref{tab:nuisance_parameters}.}

\refresponse{We carried out cosmological parameter estimation for the three marginalization methods in the 2-parameter cases, along with the ``no $n(z)$ uncertainty'' run for comparison. The contour plots in the $\Omega_m$ and $S_8$ plane are well-behaved, and the results lead to similar conclusions as for the full analyses, so we do not show them in the paper. The Figure-of-Merit ratio of the three marginalization methods to that of ``no $n(z)$ uncertainty'' is shown in green lines in Figure~\ref{fig:all_fom}, and the conclusion based on the 2-parameter cases is similar to ones drawn from the full analyses. 
}

\label{LastPage}

\end{document}